\documentclass[12pt]{article}
\pdfoutput=1
\usepackage{pstricks}
\usepackage{color}
\usepackage{cite}
\usepackage{array}
\usepackage{epsfig}
\usepackage{amssymb}
\usepackage{graphics,graphpap}
\usepackage{amssymb}
\usepackage{amsmath}
\usepackage{slashed}
\usepackage{dsfont}
\usepackage{amsfonts}
\usepackage{tikz}
\newcommand*\circled[1]{\tikz[baseline=(char.base)]{
            \node[shape=circle,draw,inner sep=2pt] (char) {#1};}}

\newcommand{\bsi}{B_6^{(1/2)}}
\newcommand{\bei}{B_8^{(3/2)}}

\def\epe{\varepsilon'/\varepsilon}

\newcommand{\RE}{{\rm Re}}

\newcommand{\gev}{\, {\rm GeV}}
\newcommand{\mev}{\, {\rm MeV}}

\newcommand{\ms}{m_{\rm s}}
\newcommand{\md}{m_{\rm d}}

\newcommand{\be}{\begin{equation}}
\newcommand{\ee}{\end{equation}}
\newcommand{\bea}{\begin{eqnarray}}
\newcommand{\eea}{\end{eqnarray}}

\newcommand{\bi}{\begin{itemize}}
\newcommand{\ei}{\end{itemize}}
\newcommand{\ord}{{\cal O}}

\def\kpn{K^+\rightarrow\pi^+\nu\bar\nu}

\def\klpn{K_{L}\rightarrow\pi^0\nu\bar\nu}

\usepackage{graphicx}

\medskipamount 

\usepackage{fancyhdr}
\advance \headheight by 3.8truept       

\newlength{\textlength}
\newlength{\overlinelength}

 \def\s#1{\setbox0=\hbox{$#1$}%
   \rlap{\ifdim\wd0>.7em\kern.22\wd0\else\kern.1\wd0\fi /}#1}


\begin{document}

\begin{titlepage}
\begin{flushright}
{FLAVOUR(267104)-ERC-60}\\
CP3-14-01\\
FERMILAB-PUB-14-001-T
\end{flushright}
\vskip0.7cm
\begin{center}
{\Large \bf \boldmath Large $N$ Approach to Kaon Decays and Mixing \\ 28 Years Later: $\Delta I=1/2$ Rule, $\hat B_K$ and $\Delta M_K$}
\vskip0.5cm
{\bf Andrzej~J.~Buras$^{a,b}$, Jean-Marc G\'erard$^{c}$ and  William A. Bardeen$^{d}$
 \\[0.4 cm]}
{\small
$^a$TUM Institute for Advanced Study, Lichtenbergstr. 2a, D-85747 Garching, Germany\\
$^b$Physik Department, Technische Universit\"at M\"unchen, James-Franck-Stra{\ss}e, \\D-85747 Garching, Germany\\
$^c$ Centre for Cosmology,
Particle Physics and Phenomenology (CP3), Universit{\'e} catholique de Louvain,
Chemin du Cyclotron 2,
B-1348 Louvain-la-Neuve, Belgium\\
$^{d}$ Fermilab, P.O. Box 500, Batavia, IL 60510, USA}
\vskip0.51cm

{\large\bf Abstract\\[10pt]} \parbox[t]{\textwidth}{\small
We review and update our results for  $K\to\pi\pi$ decays and  $K^0-\bar K^0$ mixing obtained by us in the 1980s within an {\it analytic}  approximate approach based on the dual representation of QCD as a theory of 
weakly interacting mesons for large $N$, where $N$ is the number of colours. 
 In our analytic approach the Standard Model  dynamics behind the enhancement of  ${\rm Re}A_0$ and suppression of  ${\rm Re}A_2$, the so-called
$\Delta I=1/2$ rule for $K\to\pi\pi$ decays,  has a simple 
structure: the usual octet enhancement through the long but slow quark-gluon 
 renormalization 
group evolution down to the scales $\ord(1\gev)$ is continued as a short but 
fast meson 
evolution down to zero momentum scales at which the factorization of 
hadronic matrix elements is at work. The inclusion of lowest-lying vector meson contributions in addition to the
pseudoscalar ones and of Wilson coefficients in a momentum scheme improves significantly the matching between quark-gluon 
and meson evolutions. In particular, the anomalous dimension matrix governing the meson 
evolution exhibits the structure of the known anomalous dimension matrix in the quark-gluon evolution. While this physical picture did not 
yet emerge 
from 
lattice simulations, the recent results on ${\rm Re}A_2$  and ${\rm Re}A_0$ 
from  the  RBC-UKQCD collaboration   give  support for
its correctness. 
In particular, the signs of the two main contractions found numerically by 
these authors follow uniquely from our analytic approach. 
 Though the current-current 
operators dominate the $\Delta I=1/2$ rule, working with matching scales 
$\ord(1 \gev)$ we find that the presence of QCD penguin operator $Q_6$  
 is required to obtain satisfactory result for ${\rm Re}A_0$. At NLO in $1/N$ we obtain  
$R={\rm Re}A_0/{\rm Re}A_2= 16.0\pm 1.5$ which amounts to an order of magnitude 
enhancement over the strict large $N$ limit value $\sqrt{2}$. We also 
update our results 
for the parameter $\hat B_K$, finding $\hat B_K=0.73\pm 0.02$. The smallness of $1/N$ corrections to the large $N$ value $\hat B_K=3/4$ 
results within our approach from an approximate cancellation between 
 pseudoscalar  and vector meson one-loop contributions. We 
also summarize the status of $\Delta M_K$ in this approach.
}

\vfill
\end{center}
\end{titlepage}

\setcounter{footnote}{0}

\newpage
\tableofcontents

\section{Introduction}
\label{sec:1}
Flavour violating transitions involving $K$ mesons have played a very important 
role since their discovery in 1950's, both in the construction of the Standard 
Model (SM) and more recently in the tests of its possible extensions. Unfortunately, 
due to non-perturbative uncertainties only rare $K$  decays like $\kpn$ and $\klpn$ 
can be considered as theoretically clean, that is not suffering from hadronic 
uncertainties. But here we still have to wait for sufficiently 
precise experimental results in order to see whether the SM agrees with experimental data or not.

On the other hand a number of observables in $K\to\pi\pi$ 
decays and $K^0-\bar K^0$ mixing have been measured very precisely already for quite some times. 
In quoting  their values we follow the conventions and normalizations of 
\cite{Cirigliano:2011ny}.
In particular, 
\begin{itemize}
\item
The real parts of the amplitudes $A_I$ for a kaon to decay into two pions 
with isospin $I$ are measured to be \cite{Beringer:1900zz}
\be\label{N1}
{\rm Re}A_0= 27.04(1)\times 10^{-8}~\gev, 
\quad {\rm Re}A_2= 1.210(2)   \times 10^{-8}~\gev,
\ee
and express the so-called $\Delta I=1/2$ rule \cite{GellMann:1955jx,GellMann:1957wh}
\be\label{N1a}
R=\frac{{\rm Re}A_0}{{\rm Re}A_2}=22.35.
\ee
\item
The experimental value for $K_L-K_S$ mass difference is
\be\label{DMK}
(\Delta M_{K})_{\rm exp} = 3.484(6) 10^{-15} \ \textrm{GeV}.
\ee
\item
The parameter $\varepsilon_K$, a measure of indirect CP-violation in 
$K_L\to\pi\pi$ decays, is found to be
\be\label{N2}
\varepsilon_K=2.228(11)\times 10^{-3}e^{i\phi_\varepsilon},
\ee
where $\phi_\varepsilon=43.51(5)^\circ$. 
\item
The ratio of the direct CP-violation and indirect CP-violation in $K_L\to\pi\pi$ decays 
is measured 
to be \cite{Beringer:1900zz,Batley:2002gn,AlaviHarati:2002ye,Worcester:2009qt}
\be
\RE(\epe)=1.65(26)\times 10^{-3}.
\ee
\end{itemize}

In the second half of the 1980s we have developed an approach 
to $K^0-\bar K^0$ mixing and non-leptonic $K$-meson decays 
\cite{Buras:1985yx,Bardeen:1986vp,Bardeen:1986uz,Bardeen:1986vz,Bardeen:1987vg}  based on the dual representation of QCD as a theory of 
weakly interacting mesons for large $N$, where $N$ is the number of colours 
 \cite{'tHooft:1973jz,'tHooft:1974hx,Witten:1979kh,Treiman:1986ep}.
 Reviews of our work can be found in \cite{Buras:1987vm,Bardeen:1988gu,Buras:1988kp,Buras:1988ky,Gerard:1990dx,Bardeen:1999fv,Bardeen:2001kd}. This approach provided, in particular, first results within QCD for the amplitudes ${\rm Re}A_0$ and ${\rm Re}A_2$ 
in the ballpark of experimental values. In this manner, 
for the first time, the SM dynamics behind the  $\Delta I=1/2$ rule 
has been identified. In particular, it has been emphasized that at 
scales $\ord(1\gev)$ long distance dynamics in hadronic matrix elements 
of current-current operators and not QCD-penguin operators, as proposed in 
\cite{Shifman:1975tn}, are dominantly 
responsible for 
this rule. Moreover, it has been demonstrated analytically why ${\rm Re}A_0$ is 
enhanced and why ${\rm Re}A_2$ is suppressed relative to the vacuum insertion  approximation (VIA) estimates. In this context, we have emphasized that the so-called Fierz terms in the latter approach totally misrepresent $1/N$ corrections to the strict large $N$ limit for these amplitudes.

Our approach allowed us also to calculate, for the first time within QCD, the non-perturbative parameters $\hat B_K$, $\bsi$ and 
$\bei$ 
governing the corresponding matrix elements of $\Delta S=2$ SM current-current operator and $K\to\pi\pi$ matrix elements of the 
dominant QCD-penguin ($Q_6$) and the dominant electroweak penguin ($Q_8$)
operators. Both parameters are crucial for the evaluation of $\epe$ within the SM and its various 
extensions. Also the $K\to\pi\pi\pi$ decays have been analyzed in \cite{Fajfer:1987tu} and the $K_L-K_S$ mass difference $\Delta M_K$ including long 
distance contributions has been calculated \cite{Gerard:1990dx,Bijnens:1990mz}  within this approach. During the last two decades some of these calculations 
have been improved and extended. Other applications of large $N$ ideas 
to $K\to\pi\pi$  and $\hat B_K$, but in a different spirit than our original approach, are reviewed in 
\cite{Cirigliano:2011ny}. We refer in particular to
\cite{Bijnens:1987zw,Pich:1995qp,Bijnens:1995br,Bijnens:1998ee,Hambye:1998sma,Hambye:1999ic,Peris:2000sw,Cirigliano:2002jy,Hambye:2003cy}. Recent review 
of $SU(N)$ gauge theories at large $N$ can be found in \cite{Lucini:2012gg}.

In view of the recent advances by lattice QCD on several of these parameters 
\cite{Boyle:2012ys,Blum:2011pu,Blum:2011ng,Blum:2012uk,Tarantino:2012mq,Sachrajda:2013fxa,Christ:2013lxa},
we think it is useful to improve and 
update our old results and confront them with 
the latter. We hope that  our analytic approach will shed light 
on the dynamics behind the numerical lattice computations which appear 
to indicate a pattern of long distance QCD effects in $K\to \pi\pi$ amplitudes and $K^0-\bar K^0$ mixing that is very similar to the one identified by us 
 long time ago. 

In fact, as we will discuss in more detail in the context of our presentation, 
the recent lattice results show that 
\begin{itemize}
\item
The parameter $\hat B_K$ is close to its large $N$ limit, $\hat B_K \approx 0.75$, as 
found by us in \cite{Bardeen:1987vg}.
\item
The amplitude ${\rm Re}A_2$ is suppressed through two contributions (contractions) having opposite sign and the data are reproduced within $15\%$. This pattern has been identified already in \cite{Bardeen:1986vz} and we will 
demonstrate analytically that these signs follow directly from our approach.
\item
Both in the case of ${\rm Re}A_2$ and $\hat B_K$ our 
findings of 1980s that VIA misrepresents QCD  have been recently confirmed not 
only for  ${\rm Re}A_2$ in \cite{Boyle:2012ys} but in the case of $\hat B_K$ also in \cite{Carrasco:2013jda}. This is an important confirmation as in 1988 lattice results 
provided $\hat B_K\approx 1$ \cite{Gavela:1987bd} in contradiction with the 
negative correction to the large $N$ limit for $\hat B_K$ found by us \cite{Bardeen:1987vg}. See also \cite{Gerard:2010jt} were the upper bound $\hat B_K\le0.75$ has been derived.
\item
The amplitude for ${\rm Re}A_0$ is enhanced through the contractions
 encountered in  ${\rm Re}A_2$ entering this time the amplitude with the same sign. 
In this manner another of our findings of 1980s has been confirmed. Unfortunately, as ${\rm Re}A_0$ from lattice QCD is  presently only available for non-physical kinematics, the size of this enhancement is not precisely known. Consequently a comparison between our and lattice results in this case is difficult at present.
\end{itemize}

 While according to these finding it appears that an understanding of the 
$\Delta I=1/2$ rule is emerging from lattice QCD \cite{Boyle:2012ys,Soni:2013vea}, we would like to emphasize that the suppression of  ${\rm Re}A_2$,  while 
important, is in fact
  a subleading fraction of this rule. It is  the enhancement 
of  ${\rm Re}A_0$ that is responsible dominantly for the $\Delta I=1/2$ rule.
Indeed, without short distance and long distance QCD effects $R\rightarrow\sqrt{2}$ and 
\be\label{LO}
{\rm Re}A_0\rightarrow 3.59\times 10^{-8}\gev ,\qquad   {\rm Re}A_2\rightarrow 2.54\times 10^{-8}\gev~, 
\qquad ({\rm in~large}~ N~{\rm  limit})
\ee
in plain disagreement with the data in (\ref{N1}) and (\ref{N1a}). The explanation of the  missing enhancement factor of $15.8$ in $R$ through QCD dynamics must simultaneously give the correct values for ${\rm Re}A_0$ and  ${\rm Re}A_2$. 
This means that this dynamics should suppress  ${\rm Re}A_2$ by a factor of $2.1$, not more, and enhance ${\rm Re}A_0$ by a factor of $7.5$. In our view, the understanding  of this large enhancement of ${\rm Re}A_0$ did not yet emerge from lattice QCD but has been identified at a reduced level ($5\pm1$) in our approach in 1986. We will demonstrate this in explicit terms below, improving significantly on our original estimates.

Our paper is organized as follows.  In Section~\ref{sec:0} we make a brief 
historical
review of  applications of our large $N$ framework to weak decays of mesons. 
We think this is necessary as many of the useful and important results obtained in this framework in the last thirty years appear to be unknown to younger 
generations, in particular in the lattice community. Indeed several of 
the results obtained in our papers have been confirmed in the last years 
by lattice calculations numerically
with higher control over uncertainties than it was possible in the 1980s, 
partly  due to the fact 
that the value of $\alpha_s$  was not precisely known at that time.
 In Section~\ref{sec:2} we recall the 
basic ingredients of 
the large $N$ approach to weak decays formulated in \cite{Buras:1985yx,Bardeen:1986vp,Bardeen:1986uz,Bardeen:1986vz,Bardeen:1987vg} that is based on the 
dual description of QCD at large distance scales as a truncated meson theory 
in which only pseudoscalar meson contributions were taken into account.
In Section~\ref{sec:6} we generalize this approach to 
include the effects of vector meson contributions \cite{Gerard:1988it,Gerard:1990dx}. This section is important as it 
gives further support to our approach. Indeed the inclusion of 
vector meson contributions improves 
significantly the matching between quark-gluon  
and meson pictures at scales $\ord(1\gev)$. This matching is then discussed in more detail in 
Section~\ref{Matching}.  Calculating the Wilson coefficients at NLO in a momentum scheme clarifies the relation between the 
relevant scales $M$ and $\mu$ in the effective and full theories.

It is strategically useful to illustrate our approach by discussing first 
 the $\hat B_K$ parameter. This we do  in
 Section~\ref{sec:4} including first pseudoscalar meson contributions and 
subsequently vector meson contributions.
 Armed with this technology we discuss in Section~\ref{sec:3}
the ${\rm Re}A_0$ and ${\rm Re}A_2$ amplitudes, concentrating 
on current-current operators and summarizing briefly the status of the 
parameters 
$\bsi$ and $\bei$ associated with penguin operators.
 With this information at hand we describe  in Section~\ref{sec:5a}  the understanding of the $\Delta I=1/2$ rule within our approach. We also improve and 
update the numerical analysis 
of  ${\rm Re}A_0$ and ${\rm Re}A_2$, including both current-current and 
penguin contributions. Again, the inclusion of vector meson contributions 
turns out to be important for our final results. In Section~\ref{Lattice} we compare our results from dual QCD to those available from lattice QCD. In particular the signs of various contributions found 
numerically by the  RBC-UKQCD lattice collaboration provide the confirmation of our analytic results of 1980s. Moreover, our approach allows an understanding of the
origin of these signs, which is difficult in the lattice approach.
In Section~\ref{sec:9} we focus on the $K_L-K_S$ mass difference  and
briefly mention other applications. We conclude 
in Section~\ref{sec:8}.
\boldmath
\section{Historical Review of Large $N$ Applications to Weak Decays}\label{sec:0}
\unboldmath
The first attempts to apply $1/N$ expansion to weak decays can be found in 
\cite{Fukugita:1977df,Nilles:1978ii,Tadic:1982vn}.
However, the first big step forward in the phenomenological applications of 
this expansion has been made in \cite{Buras:1985xv} in the context 
of non-leptonic charm decays, where it was realized that removing the $1/N$ 
Fierz terms from the usual vacuum insertion approximation softened the 
disagreement of the theory with both exclusive and inclusive data\footnote{This  procedure has been motivated by the analysis in \cite{Wirbel:1985ji}. However, 
these authors did not attach it with a consistent application of the $1/N$ expansion.}  This simple philosophy of using $1/N$ 
expansion has been subsequently applied to $K\to\pi\pi$ decays, $\Delta M_K$ 
and $\varepsilon_K$ in \cite{Buras:1985yx}. The 
first leading order results for the matrix elements of operators relevant for 
these
observables can be found in this paper.
Probably 
the most important results in this paper are  $\hat B_K=3/4$ \footnote{See also  \cite{Gaiser:1980gx}.} and the 
realization that the removal of $1/N$ Fierz terms from vacuum insertion calculations of current-current matrix elements suppresses ${\rm Re}A_2$, moving the theory in the direction of the data. In this paper 
also the first large $N$ result for the matrix elements of the dominant 
QCD-penguin operator $Q_6$ can be found. These leading order results 
have been subsequently confirmed in \cite{Bardeen:1986vp,Bardeen:1986uz} by 
using an effective Lagrangian describing the weak and strong interactions of 
mesons in the large $N$ limit. In particular, it has been 
emphasized in \cite{Bardeen:1986uz} that a consistent evaluation of 
the $K\to\pi\pi$ amplitudes should include two contributions:
\begin{itemize}
\item
The evolution from $M_W$ down to $\mu\approx \ord(1\gev)$, termed {\it quark-gluon 
evolution}, by means of the usual renormalization group equations. The 
result of this evolution are the values of the Wilson coefficients of 
local operators at  $\mu\approx \ord(1\gev)$.
\item
The continuation of  this evolution down to $\mu=\ord(m_\pi)$ within a meson 
theory dual to QCD, termed {\it meson evolution}. The result of this evolution 
are factorizable hadronic matrix elements.
\end{itemize}

In \cite{Bardeen:1986uz} details of quark-gluon evolution in 
the $1/N$ approach have been presented. In particular, it has been shown 
how the usual very complicated renormalization group analysis simplifies 
for large $N$, still reproducing well the exact results. In this paper 
 an additional (with respect to previous estimates) enhancement of the QCD
penguin contribution to ${\rm Re}A_0$ has been identified. It comes from 
an incomplete GIM cancellation above the charm quark mass.
In lattice calculations that work at scales $\mu=(2-3)\gev$, that are well above 
that mass,   GIM is still rather powerful
and the bulk of this contribution should be 
present in the matrix elements of current-current operators. 
Strategies 
for including charm quark contributions in lattice calculations  in the 
context of the $\Delta I=1/2$ rule and the $K_L-K_S$ mass difference
have been presented in \cite{Giusti:2004an,Endress:2012tg} and 
\cite{Christ:2012se}, respectively.

 Our studies of the 1980s culminated in the formulation of the meson evolution 
in \cite{Bardeen:1986vz,Bardeen:1987vg} and evaluation in this framework of
$1/N$ corrections to $K\to\pi\pi$ amplitudes and the parameter $\hat B_K$. These papers represent the first attempt at a consistent calculation of the weak 
matrix elements in the continuum field theory.
Pedagogical summary of this work has been presented by the authors 
in various reviews and 
lectures \cite{Buras:1988kp,Buras:1988ky,Bardeen:1988gu,Buras:1987vm,Gerard:1990dx,Bardeen:1999fv,Bardeen:2001kd}.

\boldmath
\section{Large $N$ Approach to Weak Decays of Mesons}\label{sec:2}
\unboldmath
\subsection{General Structure}\label{GeSt}
Let us begin our presentation with the general formula for the $K\to\pi\pi$ decay
amplitudes in the Standard Model \cite{Buras:1993dy}
\be\label{basic}
A(K\to\pi\pi)=\frac{G_F}{\sqrt{2}}V_{ud}V_{us}^*\sum_{i=1}^{10}(z_i(\mu)+\tau y_i(\mu))\langle \pi\pi|Q_i(\mu)|K\rangle
\ee
where 
\be
\tau=-\frac{V_{td}V_{ts}^*}{V_{ud}V_{us}^*}.
\ee
The coefficients $z_i(\mu)$ and $y_i(\mu)$ are the Wilson coefficients  
of local four-quark operators. The complete set of these operators listed 
in  \cite{Buras:1993dy}
consists of current-current 
operators  $Q_{1,2}$ , QCD-penguin operators  $Q_3-Q_6$ and 
electroweak penguin operators  $Q_7-Q_{10}$.  In our presentation only five
of them will be relevant, namely $Q_1$, $Q_2$, $Q_4$, $Q_6$ and $Q_8$. For our discussion 
it is useful to write them in the Fierz transformed form relative to 
the ones in  \cite{Buras:1993dy}. They are constructed from the light fields 
only $q=(u,d,s)$ and  are given as products of colour singlet densities, as follows
\be\label{current}
Q_1=4(\bar s_L\gamma_\mu d_L)(\bar u_L\gamma_\mu u_L), \qquad 
Q_2=4(\bar s_L\gamma_\mu u_L)(\bar u_L\gamma_\mu d_L),
\ee
\be\label{penguin}
Q_6=-8(\bar s_Lq_R)(\bar q_R d_L), \qquad Q_8=-12 e_q(\bar s_Lq_R)(\bar q_R d_L),
\ee
\be\label{Q4}
Q_4=4(\bar s_L\gamma_\mu q_L)(\bar q_L\gamma_\mu d_L)
\ee
where $q_{R(L)}=(1/2)(1\pm\gamma_5)q$ and sums over colour indices and  $q$ in $Q_4$, $Q_6$ in $Q_8$ are understood. 
Evidently, $Q_{1,2}$ are current-current operators, whereas $Q_6$ and $Q_8$ are 
the dominant density-density QCD penguin and electroweak penguin operators, 
respectively. The subleading QCD penguin operator $Q_4$ will only play a
role in our discussion of the matching of meson and quark-gluon evolutions. 
Finally, the operator relevant for $K^0-\bar K^0$ mixing 
will be  given in Section~\ref{sec:4} but the approach below applies to 
this case as well.

Since the operators $Q_i$ in (\ref{current})-(\ref{Q4}) are constructed from the light quark fields only, the full information about the heavy quark 
fields $(c,b,t)$ is contained in the Wilson coefficients $z_i$ and $y_i$. 
Correspondingly, the normalization scale $\mu$ in (\ref{basic}) is not 
completely arbitrary in our approach but must be chosen below the charm quark mass.
The values of $z_i(\mu)$ and $y_i(\mu)$ have been calculated in 1993 at the 
NLO level in the renormalization group improved perturbation theory including both QCD and QED corrections  \cite{Buras:1993dy,Ciuchini:1993vr}. Also some elements of NNLO corrections can be found in the literature \cite{Buras:1999st,Gorbahn:2004my}.

In the large $N$ approach of 
\cite{Buras:1985yx,Bardeen:1986vp,Bardeen:1986uz,Bardeen:1986vz,Bardeen:1987vg} 
the structure of different contributions to physical amplitudes is as follows.
The physics contributions from scales above $\mu$ are fully contained in 
the coefficients $z_i(\mu)$ and $y_i(\mu)$ whereas the remaining contributions 
from the low energy physics below $\mu$ (i.e. from $\mu$ to the factorization scale expected around $m_\pi$) are 
contained in the matrix elements  $\langle \pi\pi|Q_i(\mu)|K\rangle$. It follows that for $\mu=\ord(1\gev)$, the coefficients $z_i(\mu)$ and $y_i(\mu)$  can 
be calculated within a perturbative {\it quark-gluon picture} by means of renormalization group methods \cite{Buchalla:1995vs}.

As far as the meson matrix elements are concerned, the ultimate goal is to 
compute them in a nonperturbative quark-gluon picture where mesons occur as 
bound states. This route is followed by lattice computations and in fact 
since our work appeared in 1986
impressive progress has been made in this manner \cite{Tarantino:2012mq,Sachrajda:2013fxa,Christ:2013lxa}. Yet this numerical route is 
very demanding as even  
after more than a quarter of a century of hard work by lattice community
the present results for $K\to\pi\pi$ amplitudes are still 
not fully satisfactory and
the matrix element 
$\langle \pi\pi|Q_6(\mu)|K\rangle$ from lattice QCD 
is presently unknown. Moreover, it is much harder to understand the 
underlying physics than by means of an analytic approach.

Our proposal, summarized most explicitly, in \cite{Bardeen:1986vz} was to 
apply instead the
ideas of 't Hooft \cite{'tHooft:1973jz,'tHooft:1974hx} and subsequently Witten \cite{Witten:1979kh,Treiman:1986ep} to non-leptonic $K$ decays 
and $K^0-\bar K^0$ mixing. They conjectured 
that  QCD (the theory of quarks and gluons) is for large $N$ equivalent 
to a theory of weakly interacting mesons with a quartic meson coupling being 
$\ord(1/N)$.
This allows us to formulate a dual representation of the strong dynamics in terms of hadronic degrees of freedom. In the large $N$ limit, this representation becomes exact and a full description of the physics can be achieved using an infinite  set of interacting meson fields. 

The fact that QCD can be formulated both as theory of quarks and gluons on 
the one hand 
and as the theory of mesons on the other hand can now be used for Kaon mixing and non-leptonic 
decays $K\to\pi\pi$ as follows. The main point is that the matrix elements of 
four-fermion operators governing these transitions can be written at leading order in large 
$N$ as products of matrix elements of colour singlet currents in the case of current-current 
operators and as products of matrix elements of quark densities in the case 
of penguin operators. At the next-to-leading order one has two classes of 
contributions:
\begin{itemize}
\item
$1/N$ corrections to the matrix elements of factorized operators.
\item
Low energy, non-factorized matrix elements of two currents or two quark densities.
\end{itemize}

The latter contributions can be written as an integral over the momentum 
flowing through the currents (densities) in the connected planar amplitude.  One can then use our knowledge of both the high and low energy behaviour of the integrand.  At high momentum, these are just the short distance contributions to the coefficient functions of the operator product expansion which can be computed perturbatively in the quark-gluon picture. While in principle this could 
also be done in the {\it meson picture},  such an analysis would be very
 complex requiring many meson states and complicated 
interactions. However the long distance analysis is correspondingly simple as 
only lowest-lying  meson states may be required and the interactions are 
largely dictated by the chiral symmetry structure of the effective lagrangian \footnote{Moreover the spontaneous breakdown of chiral symmetry $SU(3)_L\times SU(3)_R\to SU(3)_V$ can be proven to be true in QCD in its large $N$ limit \cite{Coleman:1980mx}}. Our proposal in \cite{Bardeen:1986vz}
was to use the meson theory to interpolate to the point where one can match the behaviour of the integrand of the short distance theory.   If the amplitude is smooth enough then it may be sufficient to match the meson amplitude to the quark amplitude at an appropriate scale.   
In this manner one can achieve a consistent unified description of the physics by using the quark-gluon picture at short distances matched to the meson picture at long 
distances. The accuracy of the method depends on the interpolation of the integrand between short and long distance.

A full AdS/QCD description \cite{Witten:1998zw,Polchinski:2001tt}   should be able to interpolate the meson amplitudes to arbitrarily short distance and  first attempts in this direction have been 
made in \cite{Hambye:2005up,Hambye:2006av}.   In our approach the matching scale must presently 
be chosen around $1\gev$ implying approximate treatments in both pictures. 
In particular, the scheme dependence of the long distance part comes when one subtracts the short distance part of the integral using a particular scheme.  
This scheme dependence can be treated exactly if needed. In this context, calculating Wilson coefficients and the hadronic matrix elements in a momentum scheme we have made in the present paper a significant progress relative to our previous papers. We will discuss this important issue in  Sections~\ref{Matching} and \ref{sec:5a}.

In spite of not being exact, this approach has several virtues. Indeed,
the simplicity of this formulation lies in the fact that in the strict 
large $N$ limit QCD becomes a free theory of mesons and consequently 
the leading order contributions to any quantity are obtained by calculating 
tree diagrams with the propagated objects being mesons, 
not quarks or gluons. In this strict limit, also the factorization of 
hadronic matrix elements of four-quark operators into the product of 
matrix elements of quark currents or quark densities follows. Beyond this limit, one obtains $1/N$ expansion represented by a loop expansion in the meson theory. Even if naively these corrections could be expected to be small, one should notice 
that one-loop contributions in the meson theory represent in fact the leading 
term in the $1/N$ expansion for observables like the $\pi^+-\pi^-$ electromagnetic mass difference or
the $K^0$ decay into two neutral pions.
 In particular, 
they have to be sizable if one wants 
to explain why the subleading $K^0\to\pi^0\pi^0$ decay amplitude turns out 
to be almost equal to the $K^0\to\pi^+\pi^-$ leading one, namely the so-called 
$\Delta I=1/2$ rule.

We close this section by discussing briefly the issue of matching between 
the quark-gluon and meson theories. We will discuss this crucial issue more 
explicitly in Sections~\ref{Matching} and \ref{sec:5a}.
In the quark-gluon picture, the scale $\mu$ enters naturally as the normalization 
scale in the renormalization group improved perturbative QCD calculations 
\be\label{RGM}
\mu^2\frac{d~}{d\mu^2}Q_i(\mu^2)=-\frac{1}{2}\gamma_{ij}Q_j(\mu^2)
\ee
with $\gamma$, the anomalous dimension matrix for the $Q_i$ operators. In our 
formulation, it
serves as an infrared cut-off below which one should 
switch to the meson picture unless one wants to perform  lattice 
computations. Now the truncated meson theory, involving a finite set of light 
pseudoscalar 
and vector mesons  only,  appears non-renormalizable. In particular, 
if only lowest-lying pseudoscalar mesons are included without ultraviolet QCD completion, it exhibits a quadratic dependence 
on the cut-off which we will denote by $M$. This {\it physical} cut-off 
must be introduced in order to restrict the truncated 
meson theory to the long distance domain or, in other words, to cut-off the high mass and high momentum contributions in the meson loops. Therefore, the 
physical cut-off introduced here should be distinguished from the usual 
cut-off regularization procedure in which $M$ could be sent to arbitrarily 
large values, to disappear from observables after renormalization.

On the other hand we know that 
QCD being renormalizable has a logarithmic dependence on the ultraviolet 
cut-off. While this difference from the quadratic dependence on $M$ in 
the truncated meson theory has been in the 1980s a subject of criticism of our 
approach, one should emphasize that these two dependences are not inconsistent with 
each other. Indeed, the strict logarithmic cut-off dependence of QCD is 
valid only at short distances whereas power counting supplemented with chiral 
symmetry requires 
quadratic dependence on the cut-off for the  long distance 
behaviour  of QCD. For high values of $M$, after the inclusion of vector mesons 
and heavier meson states, this quadratic dependence on $M$ should smoothly 
turn into 
a logarithmic dependence as expected in the full meson theory. In fact, 
as demonstrated in Section~\ref{sec:6}, already the inclusion of vector mesons shows that 
this expectation is correct.

In the evaluation of the 
matrix elements $\langle \pi\pi|Q_i(\mu)|K\rangle$ the simplest choice one can 
make is $\mu=M$. This identification of $\mu$ with $M$ is certainly an idealization in the approximate treatment used in our papers, but can be improved with a complete description of quark-gluon and meson pictures used for short and long distance physics, respectively. In particular, in order to relate $\mu$ to $M$, 
one should go beyond the Fermi limit for the W-propagator and 
calculate at NLO the Wilson coefficients not in the usual 
 NDR-${\rm \overline{MS}}$ scheme but in a momentum scheme.
We will discuss this issue in Section~\ref{Matching}. Moreover, one 
should go beyond
the octet approximation for the light pseudo-scalars by including at least 
the lowest-lying vector mesons. We will do it in Section~\ref{sec:6}.

\subsection{Basic Lagrangian of the Truncated Meson Theory}
In order to calculate the hadronic matrix elements of local operators in 
our approach we use a truncated chiral Lagrangian describing the low energy 
interactions of the lightest mesons \cite{Chivukula:1986du,Bardeen:1986vp,Bardeen:1986uz}
\be\label{chL}
L_{tr}=\frac{f_\pi^2}{4}\left[Tr(D_\mu UD_\mu U^+)+rTr(m(U+U^+))-\frac{r}{\Lambda^2_\chi}Tr(m(D^2U+D^2U^+))\right]
\ee
where 
\be
U=\exp(i\frac{\Pi}{f_\pi}), \qquad \Pi=\sum_{a=1}^8\lambda^a\pi^a
\ee
is the unitary chiral matrix describing the octet of pseudoscalars. 
The singlet pseudoscalar meson $\eta_0$ decouples due to a large mass generated by 
the axial anomaly. In (\ref{chL}),  $D_\mu U$ is the usual weak covariant derivative acting on the $U$ field and $m$ is the 
real and diagonal quark mass matrix. At $\ord(p^2)$ and in the isospin limit 
$m_u=m_d= m_{ud}$,
\be
m_\pi^2= {r} m_{ud}, \quad m_K^2= \frac{r}{2}(m_s+ m_{ud}), \quad 
m_8^2=\frac{4}{3}m_K^2-\frac{1}{3}m_\pi^2.
\ee

 We would like to emphasize that the chiral Lagrangian in (\ref{chL}) must not be viewed as a normal effective tree Lagrangian but instead must be used as a fully interacting field theory including loop effects. In this sense we are providing a bosonization of the fundamental 
quark theory where all the quark currents and densities, presented below, have a valid representation in terms of the meson fields. But in the truncated version, the meson representation is valid only for a proper description of long distance physics.

The parameter $\Lambda_\chi$ in (\ref{chL}) sets the scale of higher order terms which are always expected 
in a truncated theory. It should be emphasized that this scale is a hadronic 
scale different from $\Lambda_{QCD}$. As shown in \cite{Chivukula:1986du,Bardeen:1986vz} its 
value can be determined from the physical pseudoscalar masses and decay constants:
\be
\Lambda_\chi^2= F_\pi\frac{m_K^2-m_\pi^2}{F_K-F_\pi}+\ord(\frac{1}{N}) \Rightarrow
\Lambda_\chi\approx  1.1\gev,
\ee
where we used the most recent lattice value for the ratio $F_K/F_\pi\approx 1.20$.
The $1/N$ correction, calculated in  \cite{Bardeen:1986vz}, is positive and
 in the ballpark of $5-10\%$ for the range of $M$ considered. As this correction is only logarithmically dependent on this scale, $\Lambda_\chi$ is
practicaly independent of $M$ with variation in the range $0.6\gev\le M\le 0.8\gev$ of less than $2\%$.

As stressed in \cite{Bardeen:1986vz} this cut-off independence of $\Lambda_\chi$ results only if the 
cut-off dependence of $f_\pi(M^2)$ following from our Lagrangian is taken 
into account. Explicitly one finds \cite{Bardeen:1986vz}:
\be\label{ASF}
[f_\pi^2(M^2)]^P= F_\pi^2+2 I_2(m_\pi^2)+I_2(m_K^2)
\ee
where
\be\label{I2}
I_2(m_i^2)=\frac{i}{(2\pi)^4}\int\frac{d^4q}{q^2-m_i^2}=\frac{1}{16\pi^2}\left[M^2-m_i^2\ln(1+\frac{M^2}{m_i^2})\right]
\ee
results from the calculation of one-loop diagrams in Fig.~1 of  \cite{Bardeen:1986vz}, as signalled by the $1/16\pi^2$ factor, with $M$ denoting the euclidean cut-off 
of the truncated meson theory. 
 In this manner $1/f^2_\pi(M^2)$ is the meson picture analog of the QCD running coupling in the quark picture. In fact it is amusing to note that 
$1/f^2_\pi(M^2)$ decreases with increasing $M$ implying some kind of precocious asymptotic 
freedom behaviour:
\be
\frac{\partial}{\partial M^2} \left(\frac{1}{f_\pi^2(M^2)}\right) < 0 .
\ee
With the superscript $P$ in (\ref{ASF}), we indicate that only pseudoscalar mesons have 
been included. The corresponding values of $f_\pi(M^2)$ are given in the first 
row of Table~\ref{tab:fpi}.

The chiral Lagrangian (\ref{chL}) contains only terms with a single trace 
over flavour indices which reflects the large $N$ structure of QCD. The 
leading $N$ contributions to any quantity are simply obtained from the tree 
diagrams whereas the leading $1/N$ corrections are found by calculating 
the one-loop contributions. More generally, the $1/N$ expansion corresponds to the loop expansion characterized by inverse powers of $(4\pi f_\pi)^2(f_\pi^2\sim N)$ 
with the strong interaction vertices given by the truncated Lagrangian in 
(\ref{chL}).  
It is similar to an expansion in inverse powers of $M_p^2~ (G_N = 1/M_p^2)$ if one treats general relativity as an effective field theory for gravity which is modified above the Planck scale by new degrees of freedom.\footnote{We thank John Donoghue for pointing out this analogy.}
Other details on the Lagrangian in (\ref{chL}) can be 
found in \cite{Bardeen:1986vz} and in the lecture notes \cite{Buras:1988ky,Gerard:1990dx}.

\subsection{The Structure of Hadronic Matrix Elements}
The resulting matrix elements of {\it current-current} operators 
in this approach have then the structure ($i=1,2$)
\be\label{ccm}
\langle \pi\pi|Q_i(\mu)|K\rangle=A_i\sqrt{N}\left[1+\frac{B_i(\mu)}{N}+\ord(\frac{1}{N^2})\right]
\ee
where $A_i$ and $B_i$ are $N$-independent numerical expansion coefficients 
which, in our approach, are given in terms of the parameters of the truncated 
Lagrangian. Note that the $\mu$ dependence in the matrix elements of 
$Q_{1,2}$ appears as a $1/N$ correction. This is consistent with the $\mu$ 
dependence of the Wilson coefficients $z_{1,2}(\mu)$ and reflects the simple 
fact 
that the anomalous dimensions of $Q_{1,2}$ vanish in the large $N$ limit.

On the other hand, for {\it penguin} operators $Q_6$ and $Q_8$ the matrix elements 
have the structure $(i=6,8)$
\be\label{ppm}
\langle \pi\pi|Q_i(\mu)|K\rangle=\tilde A_i(\mu)\sqrt{N}\left[1+\frac{\tilde B_i(\mu)}{N}+\ord(\frac{1}{N^2})\right].
\ee
The important difference relative to (\ref{ccm}) is the appearance of the 
$\mu$ dependence already in the leading term. Again, this is consistent with 
the $\mu$ dependence of $z_{6,8}(\mu)$ and $y_{6,8}(\mu)$ and reflects the 
fact that the anomalous dimensions of density-density operators do not 
vanish in the large $N$ limit but are twice the anomalous dimension of the mass 
operator. This fact allows a better matching of the truncated 
meson theory with the short distance contributions than it is possible for 
 the current-current operators in the case of $K\to\pi\pi$ amplitudes.

In order to calculate the matrix elements of the local operators in question we 
need meson representation of the quark currents and the quark densities. 
They are directly obtained from the effective Lagrangian in (\ref{chL}) and 
are given respectively as follows
\be\label{VAc}
\bar q^j_L\gamma_\mu q^i_L=i\frac{f_\pi^2}{4}\left\{(\partial_\mu U)U^+-U(\partial_\mu U^+)-
\frac{r}{\Lambda^2_\chi}\left[m(\partial_\mu U^+)-(\partial_\mu U)m\right]\right\}_{ij}\equiv (J_\mu)_{ij},
\ee
\be\label{RLd}
\bar q_R^j q_L^i=-\frac{f_\pi^2}{4}r\left[U-\frac{1}{\Lambda_\chi^2}\partial^2U\right]_{ij}.
\ee

We close the summary of our dual approach by stressing two major differences from  the usual chiral perturbative calculations \cite{Cirigliano:2011ny,Gasser:1983yg}:
\begin{itemize}
\item
First, the large $N$ structure of the basic truncated low energy Lagrangian 
provides a simplification over those effective Lagrangians used by chiral 
perturbation practitioners. In particular, within our ultraviolet quark-gluon 
completion, 
no $\ord(p^4)$ counter-terms are needed to absorb divergences generated by a dimensional regularization.
\item
More importantly, our loop calculations employ a cut-off regularization and 
consequently our results exhibit a {\it quadratic} dependence on the 
{\it physical} cut-off $M$. This quadratic dependence is lost in the usual 
chiral perturbative calculations, which are based on the dimensional regularization. In effect, dimensional regularization makes extra infrared subtractions of
quadratically divergent terms. These subtractions are not permitted in the 
full integration of the loop contributions in the truncated theory. As 
this quadratic dependence on the physical cut-off is usually a subject of criticism, we want to emphasize that it is an essential ingredient in the matching 
of the meson and quark-gluon pictures. Once again, it is required by 
power counting and  chiral 
symmetry. Moreover, it stabilizes the $1/N$ expansion as examplified through the 
cut-off independence of the hadronic scale $\Lambda_\chi$. Last but not least, it is at the source of the $\Delta I=1/2$ rule in our dual approach for QCD.
\end{itemize}

It is evident from these comments and from the review in  \cite{Cirigliano:2011ny} that in contrast to our $1/N$ approach, the
chiral perturbation theory framework, while being very powerful in the 
determination of low energy constants from experiment, cannot by itself 
 address the issue of the dynamics behind the $\Delta I=1/2$ rule and the evaluation of $\hat B_K$, $\bsi$ and $\bei$.

With this brief formulation of our approach at hand, we are ready to summarize the most important results obtained by us in \cite{Buras:1985yx,Bardeen:1986vp,Bardeen:1986uz,Bardeen:1986vz,Bardeen:1987vg} as well as to improve them through the inclusion of vector meson contributions (Section~\ref{sec:6})  and the calculation of Wilson coefficients in a 
momentum scheme (Section~\ref{Matching}) that allows a proper matching between meson and quark-gluon 
evolutions. Due to these improvements and the fact that  several input parameters are now much better known, our results are
more precise than in the 1980s. 
We will also  confront our findings with most recent lattice calculations in Section~\ref{Lattice}.

\begin{table}[!tb]
\centering
\begin{tabular}{|c|c|c|c|c|c|c|}
\hline
  $m_\pi$ & $m_K$ &$m_8$ & $m_V$ &  $F_\pi$   &   $F_K$  & $m_s(0.8\gev)$ \\
\hline
$135.0$ & $497.6$ & $569.3$ & $800.0$ & $91.8$   &   $110.4$  &  $155.0$ \\
\hline
\end{tabular}
\caption{Values of various quantities in units of $\mev$ discussed in the text.}
\label{tab:input}~\\[-2mm]\hrule
\end{table}

\begin{table}[!tb]
\centering
\begin{tabular}{|c|c|c|c|c|c|c|}
\hline
 $M=\mu[\gev]$  & $ 0.6$ & $0.7$ & $0.8$ & $ 0.9$&  $1.0$ & Comments \\
\hline
\hline
  \parbox[0pt][1.6em][c]{0cm}{} $f_\pi(M^2)[\mev]$ & $114.7$ & $123.5$ & $133.3$   & $143.7$ &  $154.8$ &  (P)\\
\parbox[0pt][1.6em][c]{0cm}{} $f_\pi(M^2)[\mev]$  &$107.6$  &$112.1$  &$116.4$    & $120.6$ &  $124.3$ & (P+V) \\
\hline
\end{tabular}
\caption{The anatomy of $f_\pi(M^2)$. P and V indicate that pseudoscalar and  vector mesons 
have been included or left out.
}\label{tab:fpi}~\\[-2mm]\hrule
\end{table}

\section{Inclusion of Vector Mesons}\label{sec:6}
\subsection{Preliminaries}
We will now include vector meson contributions following \cite{Gerard:1988it,Gerard:1990dx}. {As discussed in Section~\ref{GeSt}, the matrix elements of currents and densities are described by meson tree amplitudes to leading order in the 1/N expansion.  We have argued that the pion chiral Lagrangian can be used to compute the correct  infrared behavior of these amplitudes.  The vector mesons, being the next lightest states in the meson spectrum, are expected to play an important role in determining how the amplitudes evolve to higher energies.  As in deep inelastic scattering and in QCD sum rules, we expect some form of local duality to determine the interplay between neighboring states and eventually generating the smooth behavior of the perturbative short distance expansion.  By constructing an effective field theory that includes smoothly the vector meson contributions we will see how this duality begins to emerge as the amplitudes evolve in energy.  In the meson picture, additional heavy states will have to be added to continue this evolution and improve the matching further. We will return to this point below.}

In the chiral limit, the effective Lagrangian for strongly interacting pseudoscalar Goldstone bosons
\be
L(\pi)=\frac{f^2_\pi}{4} \ \mbox{Tr} \ \partial_\mu U \partial^\mu U^\dagger
\ee
is invariant under the {\it global} $SU(3)_L \times SU(3)_R$ chiral symmetry with
\be
U \to g_L U g^\dagger_R.
\ee
If we define
\be
U \equiv \xi \xi
\ee
then
\be
\xi \to g_L \xi h^\dagger (x) = h(x) \xi g^\dagger_R,
\ee
with $h(x)$ any $3\times 3$ unitary transformation, turns out to be a {\it local} symmetry of this Lagrangian.    We may thus introduce the low-lying nonet $V$ of vector mesons as the gauge bosons of this hidden $U(3)$ symmetry 
\cite{Bando:1987br} by imposing the following transformation law
\be
V \to \frac{i}{g} h \partial_\mu h^\dagger + hVh^\dagger.
\ee
In this manner, the effective Lagrangian becomes
\be\label{chiralm}
L(\pi,V) = L(\pi) -\frac{1}{4} \ \mbox{Tr} \ V_{\mu\nu} V^{\mu\nu}
-a\frac{f^2_\pi}{4} \ \mbox{Tr}\ \{\partial_\mu \xi^\dagger \xi + \partial_\mu \xi\xi^\dagger - 2ig V_\mu\}^2.
\ee
In the absence of the standard non-abelian field-strength $V_{\mu\nu}$, the vector mesons would not propagate but just be auxiliary fields such that $L(\pi,V)$ consistently reduces to $L(\pi)$ whatever the value of the free parameter $a$
associated  with the averaged $V_\mu V^\mu$ mass term. In the presence of a kinetic term for the vector mesons, they become dynamical and their low-energy properties are nicely reproduced if
$a\cong 2$.

{At this point it is useful to stress the difference between the treatment 
  of  vector bosons in our approach and in the usual 
chiral perturbation theory. It is known that in the latter approach 
 the introduction of massive spin-1 particles
(such as vector mesons) in an effective lagrangian  carries some model 
dependence.   But in the context of estimating 
the $\ord(p^4)$ chiral low-energy constants these 
 ambiguities can be removed  provided  all the models of spin-1 resonances respect  certain QCD asymptotic constraints \cite{Ecker:1989yg}.   

In our dual approach, as already stressed in the previous section, 
we do not have to worry about $\ord (p^4)$ 
chiral low-energy constants as  they are part of the quark-gluon 
evolution which satisfies the QCD asymptotic constraints . For illustration, 
in  \cite{Bardeen:1988zw}, we have explicitly shown that the "hidden-symmetry" 
approach and the "massive Yang-Mills" approach are equivalent, leading 
both to (\ref{pipimass}) for the $\pi^+-\pi^0$ mass difference discussed below.}

We will now demonstrate how the matching between the \textit{meson evolution} and the \textit{quark-gluon evolution} is significantly improved through the 
inclusion of
the nonet of light vectors in our truncated meson theory \cite{Gerard:1990dx}. In the chiral limit, they all have a  mass $m_V$ around $0.8~\gev$ and their one-loop contributions tend to transmute the quadratic cut-off dependence of 
weak hadronic matrix elements into logarithmic one.

{In this matching context, when performing the quark-gluon evolution     down to  $\mu=(0.8 - 1.0)\gev$ we should  consider all the $q\bar q$ resonances around this scale. In this spirit, the $1^{--}$     vector nonet $(\omega-\phi,\rho,K^*)$ with masses in the range $(0.77-1.02)\gev$      has to be included. The next, well-identified, $1^{+-}$ axial-vector nonet $(f_1,a_1,K_1)$ has masses in the ballpark
of $(1.23-1.43)\gev$, that is well above the matching scales we consider. It 
plays a non-negligible role in the estimate of the $\pi^+-\pi^0$ mass difference discussed below as indicated in (\ref{pipimass}). On the other hand as 
demonstrated in \cite{Gerard:1988it,Gerard:1990dx} it is by far less important for $\hat B_K$.
While it would be interesting to include these higher resonances in order 
to see the quadratic behaviour in the physical cutoff $M$ turning into a logarithmic one, from the present perspective the increased number of parameters in the corresponding effective 
Lagrangian relative to the one in (\ref{chiralm}) does not allow us to expect 
an improved precision of our approach through the inclusion of these resonances. These parameters are associated with the averaged $V_{\mu}V^{\mu}$  vector mass term, $A_{\mu}A^{\mu}$ axial-vector mass term and $A_{\mu}\partial^{\mu}\pi$ mixing  term. Future lattice simulations, if 
performed at $\ord(1\gev)$ scale, should be able to shed more light at this 
issue.}

\boldmath
\subsection{$f_\pi(M^2)$}
\unboldmath
In the chiral model defined by (\ref{chiralm}), the lowest-lying pseudoscalars are massless and the tree-level pion decay constant is modified by 
one-loop meson corrections in the following way  \cite{Gerard:1990dx}:
\be\label{FPIV}
f_\pi^2(M^2)=F_\pi^2+ \frac{3}{16\pi^2} \{(1-\frac{9a}{16})M^2 
+\frac{9a}{16}m_V^2\ln(1+\frac{M^2}{m_V^2})\}.
\ee
In the decoupling limit $m_V\gg M$, we consistently recover the quadratic 
M-dependence in  (\ref{ASF}), whatever the value of $a$. But for $a=16/9$, this 
quadratic dependence on the cut-off would totally disappear in favour of 
the logarithmic one.

Combining then equations (\ref{ASF}) and (\ref{FPIV}) with the realistic 
values $m^2_{\pi,K}\not=0$ and $a=2$ respectively, we arrive at the expression
\be\label{Fpitot}
[f_\pi^2(M^2)]^{P+V}= [f_\pi^2(M^2)]^{P}+ \Delta [f_\pi^2(M^2)]^{V}
\ee
where
\be
 \Delta [f_\pi^2(M^2)]^{V}=-\frac{27}{8} I_2(m_V^2)
\ee
with the function $I_2$ defined in (\ref{I2}). 

As seen in Table~\ref{tab:fpi} the dependence of $f_\pi$ on $M$ is now much 
weaker since  the logarithmic terms dominate now the meson evolution of 
$f_\pi(M^2)$.

Although expected from the dual representation of the strong dynamics for large $N$, such transmutation of the quadratic cut-off dependence in favour of a logarithmic one reminds us of the $\pi^+ - \pi^0$ (squared) mass splitting where a similar one-loop calculation including both the vector and the axial-vector mesons gives 
\be\label{pipimass}
\Delta m^2 (0^{-+}, 1^{--}, 1^{++}) = (\frac{3}{4 \pi}) \alpha_{\textrm{{\tiny QED}}} \int^{M^2}_0 dq^2 \frac{(m_V m_A)^2}{(q^2+m^2_V)(q^2+m^2_A)}
\ee
in agreement with the quark-gluon contribution for large   $q^2$ (i.e., $q^2 >> m^2_{V,A})$ \cite{Gerard:1990dx,Bardeen:1988zw}:
\be
\Delta m^2 ({\rm quark-gluon}) = (\frac{3}{4 \pi}) \alpha_{QED} F^2_\pi (\alpha_s r^2) \int^\infty_{M^2} \frac{dq^2}{q^4},
\ee
where $r$ is the parameter in (\ref{chL}).

In this one-loop calculation, the identification of the momentum for the virtual quarks and gluons with the momentum for the virtual mesons is straightforward since they are the same as the one carried by the color-singlet photon. So,  we are able to keep track of the momentum flow   and work in the chiral SU(2) limit for both the quark-gluon and meson evolutions. But here again, we record that the meson theory truncated to the massless pseudo-scalars leads to a pure quadratic dependence on the physical cut-off:
\be
\Delta m^2 (0^{-+}) = (\frac{3}{4 \pi})  \alpha_{QED} M^2
\ee
for small     $q^2$ (i.e., $q^2 << m^2_{V,A})$. In other words, if the ultra-violet completion for the truncated $\pi$-meson theory was not yet known, the observed $\pi^+-\pi^0$ electromagnetic mass splitting would then be explained by the existence of new degrees of freedom around the cut-off $M \approx 0.85$ GeV.

 \boldmath
\subsection{$J_\mu\otimes J^\mu(M^2)$}
\unboldmath
A similar though not so striking transmutation occurs for left-handed current-current operators in the chiral model defined in (\ref{chiralm}):

\begin{align}\begin{split}\label{JJ}
\{(J_\mu)_{ij}(J^\mu)_{kl}\}(M^2)= &\{(\bar J_\mu)_{ij}(\bar J^\mu)_{kl}\}(0)-c(M^2)\{(J_\mu)_{il}(J^\mu)_{kj}\\ & -
\frac{1}{2}[\delta_{il}(J_\mu J^\mu)_{kj}+\delta_{kj}(J_\mu J^\mu)_{il}]\}(0)
\end{split}
\end{align}
with $(J_\mu)_{ij}$ defined in (\ref{VAc}) and
\be 
 \bar J_\mu=i \frac{F_{\pi,K}}{2} \partial_\mu\pi+\frac{i}{4}[(\partial_\mu \pi) \pi-\pi (\partial_\mu\pi)] + ......
\ee
 being the relevant $\Delta S=0,\pm 1$ physical hadronic current. Moreover
\be\label{cM2}
c(M^2)=\frac{1}{16\pi^2}\left[\frac{2 M^2}{f^2}\right]+\frac{3a}{16 f^2}\{(a-5)I_2(m_V^2)+a m_V^2 I_3(m_V^2)\},
\ee
where
the function $I_3(m_i^2)$ is just the derivative of $I_2(m_i^2)$ with respect to $m_i^2$
\be\label{I3}
I_3(m_i^2)\equiv\frac{dI_2(m_i^2)}{dm_i^2}=\frac{1}{16\pi^2}\left[\frac{M^2}{M^2+m_i^2}-\ln(1+\frac{M^2}{m_i^2})\right].
\ee
For ($m_V\to\infty$, $a$ arbitrary) and ($a\to 0$, $m_V$ arbitrary) only the 
first term on the r.h.s of  (\ref{cM2}) survives, corresponding precisely to the pseudoscalar contribution in the chiral limit.

\section{Matching of Meson and Quark-Gluon Evolutions}\label{Matching}
The identification of $M$ and $\mu$ has been the subject of  criticism in 
the past. Therefore we would like to discuss this point and present an improved 
treatment that goes beyond our work of 1980s. 
First, as discussed in particular in \cite{Bardeen:1988gu,Bardeen:1999fv,Bardeen:2001kd} and  \cite{Fatelo:1994qh}, in the large $N$ expansion the nonfactorized 
amplitudes responsible for both meson and quark-gluon evolutions are given by 
a convolution of the $W$-boson propagator $D^{\mu\nu}_W(q)$ with a  tree amplitude $A_{\mu\nu}$ as follows
\be\label{NFA}
A(p_1,..p_n)=i\int\frac{d^4q}{(2\pi)^4}D^{\mu\nu}_W(q)A_{\mu\nu}(q,p_1,..p_n).
\ee
The short and long distance contributions to this amplitude are controlled by 
the explicit momentum flowing through the $W$-boson propagator. These contributions can be separated by a suitable regularization of this integration. An explicit example is provided by the analytic regularization
\be\label{analytic}
D^{\mu\nu}_W(q)\to D^{\mu\nu}_W(q)\left[\frac{q^2}{q^2-M^2}+\frac{-M^2}{q^2-M^2}\right],
\ee
which we will use in what follows.
The first term contributes at short distances but is suppressed at low momentum.  The second term contributes at long distances but the high momentum components 
are suppressed. This separation can be exploited to use the quark-gluon representation for the first term and a truncated meson Lagrangian for the 
second term. 

Now, it is well known that the Wilson coefficients depend on renormalization 
scheme and are usually computed using dimensional regularization for UV and various schemes for $\gamma_5$ in $D$ dimensions like NDR or HV schemes \cite{Buras:1998raa}. In order 
to be able to make the identification 
\be\label{scales}
(M^2)_{\textrm{{\scriptsize mesons}}}    = (\mu^2)_{\textrm{{\scriptsize quark-gluon}}}, \ee
we have to relate
the Wilson coefficients calculated in these schemes to the ones in which the 
integral in (\ref{NFA}) is calculated in $D=4$ with an UV momentum cut-off. We 
call this scheme ${\rm \overline{MOM}}$ scheme. The {\it bar} indicates that 
this scheme should not be confused with momentum schemes used in the past 
for short distance calculations.

This shift in Wilson coefficients can be found 
as usual by calculating perturbatively in the quark-gluon picture one-loop
matrix elements of operators in different schemes for UV but using the same IR 
regulator and comparing the finite non-logarithmic pieces. In the case at hand, 
retaining only the first term in (\ref{analytic}) and setting the external 
momenta to be zero corresponds effectively to regulating IR divergences by 
giving the mass $M$ to the gluon. Proceeding in this manner the
coefficients $z_1$ and $z_2$ in the ${\rm \overline{MOM}}$ scheme to be combined with the meson evolution can be obtained from the known coefficients  calculated 
in the  NDR-${\rm \overline{MS}}$ scheme. Details can be found in 
\cite{Bardeen:2001kd}, where the same results for the ${\rm \overline{MOM}}$ scheme 
 have been obtained calculating the shift relative to the HV scheme. 
One finds then 
\be\label{z1}
z_1({\rm \overline{MOM}})= z_1({\rm NDR})+\frac{\alpha_s}{4\pi} \frac{11}{2N}z_1({\rm NDR})-\frac{\alpha_s}{4\pi} \frac{11}{2}z_2({\rm NDR}),
\ee
\be\label{z2}
z_2({\rm \overline{MOM}})= z_2({\rm NDR})-\frac{\alpha_s}{4\pi} \frac{11}{2}z_1({\rm NDR})+\frac{\alpha_s}{4\pi} \frac{11}{2N}z_2({\rm NDR}).
\ee
These results have been confirmed in the present paper.

As the matrix elements of $Q_6$ operator are presently known only in 
the large $N$ limit, it is sufficient to use for $z_6$ coefficient 
only its LO result that is in any case  $\ord(1/N)$ and moreover  GIM 
suppressed. To this 
end we will use  $\alpha_s$ given in the $\overline{\rm MS}$ scheme and 
the known leading order anomalous dimensions.

The results for $z_1$, $z_2$  in ${\rm \overline{MOM}}$ and NDR schemes for different $\mu$ 
are given in Table~\ref{tab:Results2}. We observe large enhancement of $|z_{1,2}|$ in the momentum scheme over the values in the NDR scheme. We have checked 
that for the same value of the coupling constant the leading order values of $z_{1,2}$ are between those obtained in ${\rm \overline{MOM}}$ and NDR schemes. 

\begin{table}[!tb]
\centering
\begin{tabular}{|c|c|c|c|c|c|c|}
\hline
 $\mu[\gev]$  & $ 0.6$ & $0.7$ & $0.8$ & $ 0.9$&  $1.0$ & Comments \\
\hline
\hline
\parbox[0pt][1.6em][c]{0cm}{}$\alpha_s(\mu)$ &$0.812$ &$0.658$ &$0.564$& $0.502$&
$0.457$ &  ${\rm \overline{MS}}$   \\
\parbox[0pt][1.6em][c]{0cm}{}$z_1(\mu)$ & $-1.228$  & $-1.029$ & $-0.900$&$-0.809$&$-0.740$& ${\rm \overline{MOM}}$  \\
 \parbox[0pt][1.6em][c]{0cm}{}$z_2(\mu)$ & $1.777$  & $1.625$& $1.530$ &$1.463$& $1.415$&  ${\rm \overline{MOM}}$ \\
\parbox[0pt][1.6em][c]{0cm}{}$z_6(\mu)$ & $-0.069$  &$-0.049$& $-0.037$ &           $-0.029$& $-0.023$    & ${\rm LO}$\\
\parbox[0pt][1.6em][c]{0cm}{}$z_1(\mu)$ & $-0.660$  & $-0.590$ & $-0.537$&$-0.495$&$-0.461$& NDR-${\rm \overline{MS}}$  \\
 \parbox[0pt][1.6em][c]{0cm}{}$z_2(\mu)$ & $1.379$  & $1.328$& $1.291$ &$1.262$& $1.240$&NDR-${\rm \overline{MS}}$ \\
\parbox[0pt][1.6em][c]{0cm}{}$z_6(\mu)$ & $-0.097$  &$-0.065$& $-0.047$ &           $-0.035$& $-0.027$    & NDR-${\rm \overline{MS}}$ \\
\hline
\end{tabular}
\caption{Values of the Wilson coefficients $z_i$  as functions 
of $\mu$ for the ${\rm \overline{MOM}}$ and NDR-${\rm \overline{MS}}$ schemes.  
}\label{tab:Results2}~\\[-2mm]\hrule
\end{table}

In order to complete the matching we have to calculate the relevant loop 
diagrams in the meson theory, including in the integrands the second term 
in (\ref{analytic}). We find then a simple rule for transforming the results of our previous papers into the ones obtained using an analytic regularization that can be properly combined with the coefficients $z_i$ in ${\rm \overline{MOM}}$ scheme. One just has to 
replace
the function $I_2(m_i^2)$ in (\ref{I2}) by\footnote{Needless to say this 
replacement should not be made in the calculation of $f_\pi$, where no 
meson evolution is involved.}
\be\label{hatI2}
\hat I_2(m_i^2)=\frac{i}{(2\pi)^4}\int\frac{d^4q}{q^2-m_i^2}\left[\frac{-M^2}{q^2-M^2}\right]=\frac{1}{16\pi^2}\frac{M^2}{M^2-m_i^2}\left[\ln(2) M^2-m_i^2\ln(1+\frac{M^2}{m_i^2})\right]
\ee
with the limiting value for $M^2=m_V^2$
\be
\hat I_2(m_V^2)=\frac{1}{16\pi^2} m_V^2\left[\ln(2)-\frac{1}{2}\right].
\ee

Similarly, the function $I_3(m_i^2)$ in (\ref{I3}) should be replaced  
by the  derivative of $\hat I_2(m_i^2)$ in (\ref{hatI2}). We find then 
\be\label{hatI3}
\hat I_3(m_i^2)=\frac{1}{16\pi^2}\frac{M^4}{(M^2-m_i^2)^2}\left[\ln(2)+
\frac{M^2-m_i^2}{M^2+m_i^2}-\ln(1+\frac{M^2}{m_i^2})\right]
\ee
and the limiting value for $M^2=m_V^2$
\be
\hat I_3(m_V^2) =-\frac{1}{16\pi^2} \left[\frac{1}{8}\right].
\ee

It should be noted that the presence of $\ln(2)$ multiplying $M^2$ will in 
turn decrease in the ${\rm \overline{MOM}}$ scheme the matrix elements relative to our 
previous results, while as we have seen above in the ${\rm \overline{MOM}}$ scheme $|z_{1,2}|$ 
become consistently larger than in the LO.

\boldmath
\section{The Parameter $\hat B_K$}\label{sec:4}
\unboldmath
\subsection{Preliminaries}
As the physics behind the 
$\Delta I=1/2$ rule is more involved than the one in $K^0-\bar K^0$ mixing, 
it is strategically useful to apply first our approach to the calculation of 
the parameter $\hat B_K$. We will  first only 
include pseudoscalar meson contributions, but subsequently also vector meson 
contributions will be taken into account. This will demonstrate explicitly that the inclusion of vector mesons significantly improves the matching between the meson and quark-gluon pictures.

The renormalization group invariant 
parameter is given as follows \cite{Buras:1990fn}
\be\label{BKhat}
\hat B_K=B_K(\mu)\left[\alpha_s^{(3)}(\mu)\right]^{-b}
\left[1+\frac{\alpha_s^{(3)}(\mu)}{4\pi}J_3\right], \qquad b=\frac{9(N-1)}{N(11N-6)},
\ee
where we have shown the $N$-dependence of the exponent $b$ in the leading term to 
signal that $b$ vanishes in the large $N$ limit. The coefficient $J_3$ is renormalization scheme dependent. It has been calculated in the  NDR-${\rm \overline{MS}}$ in \cite{Buras:1990fn}. However, as discussed in the previous section, in our approach we 
have to work in a ${\rm \overline{MOM}}$ scheme. 

As the operator $\Delta S=2$ and the $\Delta I=3/2$ operator have the same 
anomalous dimension, relations in (\ref{z1}) and (\ref{z2}) allow us 
to calculate the shift in $J_3$ entering $\hat B_K$ in (\ref{BKhat}). From 
the $\ord(\alpha_s)$ term in the sum $z_1+z_2$ we obtain
\be
J_3({\rm \overline{MOM}})=J_3({\rm NDR})-\frac{11}{2}(1-\frac{1}{N}).
\ee
Using the known NDR result from \cite{Buras:1990fn} and setting $N=3$ we find 
\be\label{J3NDRMOM}
J_3=1.895~{\rm (NDR)}, \qquad J_3=-1.772~{\rm (\overline{MOM})}.
\ee

 The scale dependent parameters $B_K(\mu)$ is related 
to the relevant hadronic matrix element of the $\Delta S=2$ operator
\be\label{DS2OP}
Q(\Delta S=2)=4(\bar s_L\gamma^\mu d_L)(\bar s_L\gamma_\mu d_L)
\ee
as follows
\be
\langle \bar K^0|Q(\mu)|K^0\rangle = B_K(\mu)\frac{16}{3}F_K^2 m_K^2.
\ee

The normalization of $B_K$ is such that in the vacuum insertion approximation 
$B_K$ is unity. Indeed 
\be\label{vacuum}
B_K(\mu)=\frac{3}{4}\left(1+\frac{1}{N}\right) =1, \qquad {\rm (in~VIA)}
\ee
where the $1/N$ represents again the Fierz-term. As already 
stressed in  \cite{Bardeen:1987vg},
this term completely misrepresents the full $1/N$ correction to the leading term. 
Its positive sign  as opposed to the negative sign required for the 
matching with $1/N$ corrections in the quark-gluon evolution and 
 the absence of any $\mu$ dependence in this result show that 
it is incompatible with the quark-gluon picture of QCD. This has been 
recently confirmed in lattice QCD \cite{Carrasco:2013jda}.

On the other hand, the leading term of $B_K=3/4$ \cite{Buras:1985yx,Gaiser:1980gx} 
is the correct prediction of  truncated meson theory in the 
strict large $N$ limit. Indeed, 
at this stage the following important point should be made. As in the strict 
large $N$ limit the exponent in (\ref{BKhat}) and the NLO term involving 
$J_3$ vanish, we find that 
independently of any renormalization scale or renormalization scheme for 
the operator $Q(\Delta S=2)$ in the 
large $N$ limit
\be\label{BKLO}
\hat B_K \rightarrow 0.75, \qquad ({\rm in~large~N~limit}).
\ee
The question then arises whether after the inclusion of $1/N$ corrections 
$\hat B_K$ is larger or smaller than its leading value. In the 1980s 
the values of $\hat B_K$ varied from $1/3$ obtained using PCAC-SU(3) \cite{Donoghue:1982cq}, 
$0.40$ obtained through hadronic sum rules \cite{Pich:1985ab} to values close to unity, 
obtained in particular within the lattice approach \cite{Gavela:1987bd}. On the 
other hand, using the truncated meson theory 
outlined in Section~\ref{sec:2} and thereby including only pseudoscalar meson 
contributions we have found  $1/N$ corrections to be small
and 
{\it negative}. However, the left-over albeit weak $\mu$ dependence of $\hat B_K$ and 
the inaccurate value of $\alpha_s$ at that time 
 lead us to a rather conservative error on $\hat B_K$  \cite{Bardeen:1987vg}
\be\label{BK1987}
\hat B_K=0.66\pm 0.07, \qquad  ({\rm in~dual~QCD,}~1987).
\ee
Shortly after, it has been shown that the  inclusion of vector mesons in this calculations \cite{Gerard:1988it,Gerard:1990dx}
moved $\hat B_K$ much closer to its leading order value in (\ref{BKLO}).
Since then several semi-analytic calculations by other authors have been performed. They are reviewed in \cite{Cirigliano:2011ny}.

On the other hand, a quarter of century after our first result the world lattice average for 
$\hat B_K$ based on the calculations of various groups \cite{Aoki:2010pe,Bae:2010ki,Constantinou:2010qv,Colangelo:2010et,Bailey:2012bh,Durr:2011ap} 
reads for $N_f=2+1$ calculations (recent FLAG update of \cite{Colangelo:2010et})
\be\label{L2012}
\hat B_K=0.766\pm 0.010, \qquad   ({\rm in~ lattice~QCD,}~2013).
\ee
See also the very recent analyses in \cite{Bae:2013lja,Frison:2013fga,Bae:2014sja}.
 The following remarks are in order
\begin{itemize}
\item
The precision of lattice result is truly impressive, though on the verge of 
being challenged by isospin breaking effects.
\item
The value in (\ref{L2012}) is consistent with our estimate in (\ref{BK1987}) 
 which we will update below. Moreover it
is very close to the leading $N$ result in (\ref{BKLO}).
\item
The sign of the correction to leading $N$ result 
obtained presently in the lattice calculations
appears to be {\it positive} and not {\it negative} as favoured in our 
framework and discussed below.
\end{itemize}

After recalling the analytic expressions for $\hat B_K$ in the truncated 
meson theory and including vector meson contributions, we will give arguments in favour of a {\it negative} 
correction to the leading large $N$ result so that in QCD we expect:
\be\label{Bound}
\hat B_K\le 0.75~,\qquad  ({\rm in~1/N~ expansion}).
\ee
Therefore we believe that the lattice error in (\ref{L2012}) is underestimated 
and the improved lattice calculations will satisfy the bound in (\ref{Bound}) 
giving values for $\hat B_K$ a bit lower than the present world lattice 
average. In fact, a number of lattice groups among \cite{Aoki:2010pe,Bae:2010ki,Constantinou:2010qv,Colangelo:2010et,Bailey:2012bh,Durr:2011ap} published results with central 
values satisfying the bound in (\ref{Bound}) but the errors did not allow 
for a clear cut conclusion.

\boldmath
\subsection{Calculating $\hat B_K$ in the Truncated Meson Theory}
\unboldmath
Including one-loop 
contributions in the meson theory truncated to pseudoscalar mesons as done in 
 \cite{Bardeen:1987vg} and performing the replacements (\ref{hatI2}) and (\ref{hatI3}), we find
\be\label{BKF}
B^P_K(M)= \frac{3}{4}\left\{1-\frac{1}{4F^2_K}\left[3(1+\frac{m_8^2}{m_K^2})\hat I_2(m_8^2)+(1+\frac{m_\pi^2}{m_K^2})\hat I_2(m_\pi^2)+4 m_K^2 \hat I_3(m_K^2)\right]\right\}.
\ee
 With the superscript $P$ we indicate that only pseudoscalar mesons have 
been included.

In Table~\ref{tab:BKResults} we give $B_K(\mu)$ and $\hat B_K$ obtained using (\ref{BKhat}) 
and (\ref{BKF}) with $\mu=M$. We confirm that while  $B_K(\mu)$ depends strongly
on $\mu$, this $\mu$ dependence is cancelled significantly by the 
$\mu$ dependent factor coming from the QCD analysis in the quark-gluon picture. 
On a semi-quantitative level this reduction of $\mu$ dependence  shows that the quark-gluon and the meson pictures of strong interactions match well as required for the consistency of our calculation. Yet, for $M=(0.6-0.7)\gev$ the accuracy 
of the $\mu$ dependent factor coming from the quark-gluon picture cannot be 
trusted as in this range $\alpha_s(\mu)\ge 0.65$. On the other 
hand,  for $M\ge0.8\gev$, 
$\hat B_K$ shows a significant $M$ dependence signalling that the meson 
evolution described by means of pseudoscalars only ceases to be a good 
approximation. Therefore in order to decrease the gap between the validity 
of both pictures, the inclusion of vector mesons is necessary. Yet, already 
at this stage we note two facts
\begin{itemize}
\item
$B^P_K(M)$ given here in the ${\rm \overline{MOM}}$ scheme differs significantly from the values 
quoted by lattice groups that use the NDR-${\rm \overline{MS}}$ scheme. In 
the latter scheme the values of this parameter are much lower.
\item
However, this difference is compensated by the  
QCD factor in  (\ref{BKhat}) being significantly above unity in the NDR-${\rm \overline{MS}}$ scheme, while it is  close to unity in the ${\rm \overline{MOM}}$ scheme. Indeed the LO 
enhancement of this factor present in any scheme is significantly compensated by the negative NLO correction in the ${\rm \overline{MOM}}$ scheme as seen in (\ref{J3NDRMOM}). 
\end{itemize}

It should be noted that in the chiral limit, $m_{\pi,K}^2 \to 0$, the 
result in (\ref{BKF}) implies 
\be\label{chiral}
B_K(M) = \frac{3}{4} \left(1-\frac{2M^2}{(4\pi F_K)^2} \right),
\ee
so that for $M=0.7\gev$ one finds $B_K(M)=0.37$. As seen in 
  Table~\ref{tab:BKResults} this strong suppression is significantly 
softened for $m_{\pi,K}^2\not=0$, see also \cite{Bijnens:1995br}, but 
on the whole the resulting value of $\hat B_K$ is visibly below the 
lattice value in (\ref{L2012}). 
As we 
will now demonstrate, after the inclusion of vector contributions, the final result 
for $\hat B_K$ will turn out to be very close to the lattice result.

\begin{table}[!tb]
\centering
\begin{tabular}{|c|c|c|c|c|c|c|}
\hline
 $M[\gev]$  & $ 0.6$ & $0.7$ & $0.8$ & $ 0.9$&  $1.0$ & {\rm Comments}\\
\hline
\hline
\parbox[0pt][1.6em][c]{0cm}{}$B^P_K(M)$&$0.698$  &$0.665$& $0.622$ &$0.568$& $0.502$& (P)\\
\parbox[0pt][1.6em][c]{0cm}{}$\hat B^P_K$& $0.647$  & $0.662$&$0.650$ 
& $0.615$ &$0.559$ &\\
\hline
\parbox[0pt][1.6em][c]{0cm}{}$B_K(M)$ &$0.728$  &$0.716$& $0.700$ &$0.679$&
$0.653$ & (P+V)\\
\parbox[0pt][1.6em][c]{0cm}{}$\hat B_K$ &$0.676$  &$0.713$& $0.731$& $0.735$&$0.728$ & \\
 \hline
\end{tabular}
\caption{The anatomy of $B_K$ as function 
of the scale $M$. 
}\label{tab:BKResults}~\\[-2mm]\hrule
\end{table}

\boldmath
\subsection{Inclusion of Vector Meson Contributions in $\hat B_K$}
\unboldmath
From the generic formula (\ref{JJ}), one easily infers how
the inclusion of the lowest-lying  vector mesons modifies the  cut-off dependence of the 
$B_K$ parameter. In the chiral limit and 
for $a=2$ one has \cite{Gerard:1988it,Gerard:1990dx}
\be\label{BKV}
B_K(M) = \frac{3}{4} \left\{1-\frac{1}{(4\pi F_K)^2} \left[\frac{7}{8}M^2
+\frac{3}{8}m^2_V \ln (1+\frac{M^2}{m^2_V})
+\frac{3}{4}\frac{m^2_VM^2}{(M^2+m^2_V)}\right]\right\} .
\ee
In the decoupling limit $m_V\gg M$, we consistently recover the result in 
(\ref{chiral}).
But for  $M> m_V$, we observe a reduction by more than $50\%$ of the quadratic dependence on the cut-off. 
Once again, this transmutation of the quadratic cut-off dependence in favour of a logarithmic one with the same sign is clearly linked to the introduction of a new intrinsic scale $m_V$ which changes the power counting in a way still consistent with chiral symmetry. 

Again, as in the case of pseudoscalar contributions, we have to adjust the 
result in (\ref{BKV}) to the ${\rm \overline{MOM}}$ scheme by means of the procedure summarized in 
Section~\ref{Matching}.
Combining then equations (\ref{BKF}) and (\ref{BKV}) properly modified by this 
procedure and taking into account
that the contribution (\ref{chiral}) is already present in (\ref{BKF}), we 
arrive at the expression
\be\label{BKtot}
B^{\rm tot}_K(M)= B^P_K(M)+ \Delta B_K^V(M)
\ee
where
\be
 \Delta B_K^V(M)=\frac{3}{(4 F_K)^2}\left[\frac{9}{2} \hat I_2(m_V^2)-3 m_V^2\hat I_3(m_V^2)\right]
\ee
with the functions $\hat I_2$ and $\hat I_3$ defined in (\ref{hatI2}) and (\ref{hatI3}), 
respectively. 

In Table~\ref{tab:BKResults} we show the results obtained using (\ref{BKtot}). 
The effect of the reduction of 
$\mu$ dependence in $\hat B_K$ is very significant when compared with the 
pseudoscalar case, again demonstrating that our evolution picture is correct.
This is in particular the case in the range $M=(0.7-0.9)\gev$ where 
we expect our truncated meson theory after the inclusion of vector mesons 
to give reliable results. 

We note that the effect of inclusion of vector meson has only a small impact 
at $M=0.6\gev$ but this impact increases quickly with increasing $M$. In 
particular, the value of $\hat B_K$ is increased and turns out to be close to its leading value as 
the vector meson contributions enter with the opposite sign to the 
pseudoscalar meson contributions.
On the basis of these 
results 
we quote our final result
\be\label{BKfinal}
\hat B_K= 0.73\pm 0.02,
\ee
where the error should not be considered as a standard deviation. Rather, 
this result represents the range for $\hat B_K$ we expect in our approach. 
The lower value corresponds to the value at $M=0.7\gev$ which should 
be sufficiently large so that our calculation is reliable and the upper bound 
is  just the bound in (\ref{Bound}) to which we will return below. We 
consider this range as conservative as the $M$ dependence of  $\hat B_K$
 displayed in 
 Table~\ref{tab:BKResults} amounts
for $M=(0.8-1.0)\gev$  to only $1\%$.

This result is in an excellent agreement with the lattice QCD value 
in (\ref{L2012}) although we are aware of the fact that while lattice 
calculations have good control over their errors, this is not quite 
the case here. Still it is encouraging that such a simple analytic 
approach could provide the explanation why the lattice results turn out 
to be so close to the strict large $N$ limit value of $\hat B_K$.

In summary, we observe that within our approach the smallness of $1/N$ corrections to the leading result for $\hat B_K$ follows from an approximate cancellation between pseudoscalar  and vector meson one-loop contributions. Moreover,  this cancellation is consistent with the small anomalous dimension of the 
$\Delta S=2$ operator and consequently allows a good matching of meson and 
quark-gluon evolutions.

{Finally, we would like to refer to the analysis in \cite{Bijnens:1995br} 
    which was done in the spirit of our approach except that for the 
  low energy meson contributions an Extended Nambu-Jona-Lasinio model has 
  been used. Moreover, a sharper matching between long distance and short distance contributions has been performed at the LO level in $\alpha_s$. 
The result 
 $0.60\le\hat B_K\le 0.80$, even if less precise, is fully consistent 
with the values obtained in our approach.}

\boldmath
\subsection{An Upper Bound on $\hat B_K$}
\unboldmath
Let us next discuss the sign of $1/N$ corrections to the leading 
result in (\ref{BKLO}). 
In fact, the existence of the upper bound on the $\hat B_K$ parameter in (\ref{Bound}) has been demonstrated in \cite{Gerard:2010jt} and we recall briefly the 
main arguments here. To derive this bound, let us exchange a fictitious color-singlet boson between the two left-handed currents of the $\Delta S = 2$ operator  in (\ref{DS2OP}). In the $1/N$ expansion, the full leading and next-to-leading contributions to $B_K$ can then be viewed as two-bubble and one-bubble topologies, respectively (see Fig.~2 of \cite{Gerard:2010jt}). In this simple pictorial approach, the $1/N$ Fierz-term is clearly part of the second disconnected topology. For each closed quark loop (wherein the sum over all planar gluons is understood), we take indeed the trace over colours. But for each closed fermion loop, we also have to multiply by the spin-statistics factor $(-1)$. This results in a negative $1/N$ correction to the leading value of the $B_K$ parameter.

As seen in Table~\ref{tab:BKResults} our results for $\hat B_K$ satisfy the
 upper bound in question. On the other hand, the central value of  $\hat B_K$ from lattice simulations in (\ref{L2012}) violates this bound but is consistent 
within $2\sigma$. We expect therefore that improved lattice calculations 
will satisfy our  bound one day and in a few years from now lattice researchers will 
quote  $\hat B_K\approx 0.74$. In fact, the most recent update from staggered 
quarks \cite{Bae:2013lja,Bae:2014sja} quotes precisely $\hat B_K=0.738\pm0.005$ but 
additional systematic error of $0.037$ does not allow for definite conclusions.

\boldmath
\section{${\rm Re}A_0$ and ${\rm Re}A_2$ Amplitudes}\label{sec:3}
\unboldmath
\subsection{Preliminaries}
The amplitudes for $K\to\pi\pi$, neglecting the $\Delta I=5/2$ contributions, can be parametrized in terms of
isospin amplitudes $A_I$ through \cite{Cirigliano:2011ny}
\begin{equation}  
A(K^+\rightarrow\pi^+\pi^0)=\frac{3}{2} A_2 e^{i\delta_2}
\end{equation}
\begin{equation} 
A(K^0\rightarrow\pi^+\pi^-)=A_0 e^{i\delta_0}+ \sqrt{\frac{1}{2}} A_2 e^{i\delta_2}
\end{equation}
\begin{equation}
A(K^0\rightarrow\pi^0\pi^0)= A_0 e^{i\delta_0}-\sqrt{2} A_2 e^{i\delta_2}\,.
\end{equation} 
Here the subscript $I=0,2$ denotes states with isospin $0,2$
equivalent to $\Delta I=1/2$ and $\Delta I = 3/2$ transitions,
respectively, and $\delta_{0,2}$ are the corresponding strong phases. 
The weak CKM phases are contained in $A_0$ and $A_2$. The experimental values 
of these amplitudes are given in the isospin limit in (\ref{N1}).
The strong phases $\delta_{0,2}$ cannot be calculated in our framework since the  $\pi\pi$ elastic rescattering has no ultraviolet completion. 
Their difference is measured to be \cite{Beringer:1900zz}
\be
\delta_0-\delta_2=(47.5\pm0.9)^\circ~.
\ee

Equivalently, we have
\be\label{DI1}
 A_0 e^{i\delta_0}=\frac{1}{3}\left[2 A(K^0\rightarrow\pi^+\pi^-)+
A(K^0\rightarrow\pi^0\pi^0)\right],
\ee

\be\label{DI2}
 A_2 e^{i\delta_2}=\frac{\sqrt{2}}{3}\left[A(K^0\rightarrow\pi^+\pi^-)-
A(K^0\rightarrow\pi^0\pi^0)\right],
\ee
where we use  the following isospin relation 
\be\label{DI3}
A(K^0\rightarrow\pi^+\pi^-)-A(K^0\rightarrow\pi^0\pi^0)=\sqrt{2}A(K^+\rightarrow\pi^+\pi^0)
\ee
which
provides a consistency check when extracting all non-vanishing hadronic 
matrix elements.

\subsection{Meson Evolution of Current-Current Operators}
In the limit $m^2_\pi \to 0$, the four $K\to\pi\pi$ one-loop diagrams given in Fig.~2 of \cite{Bardeen:1986vz} can be viewed as a meson operator evolution down to the factorization scale:
\be\label{ME1}
Q_1(M^2)=Q_1(0)-c_1(M^2) Q_2(0)
\ee

\be\label{ME2}
Q_2(M^2)=Q_2(0)-c_1(M^2) Q_1(0) +c_2(M^2)[Q_2(0)-Q_1(0)].
\ee
with {\it positive} coefficients
\be\label{c1P}
c_1(M^2)\approx \frac{1}{(4\pi f_\pi)^2}\left[\frac{f_\pi}{F_\pi}\right]
  \left\{2 \hat M^2-\frac{m_K^2}{4}\ln(1+\frac{M^2}{\tilde m^2})\right\},
\ee

\be\label{c2P}
c_2(M^2)\approx \frac{1}{(4\pi f_\pi)^2}\left[\frac{f_\pi}{F_\pi}\right]
  \left\{ \hat M^2+ {m_K^2}\ln(1+\frac{M^2}{\tilde m^2})\right\},
\ee
where the $M^2$ dependence of the expansion parameter $f_\pi$, given in (\ref{ASF}), has not been written explicitly. These evolution equations, the positivity of the coefficients 
$c_i$ and the fact that $c_i=\ord(1/N)$ are fundamental for our explanation of
 the $\Delta I=1/2$ rule. It originates in the continuation of the usual quark-gluon evolution by means of meson evolution below scales $\ord(1)\gev$ down to 
factorization scale at which QCD becomes a theory of free interacting mesons. 
In what follows we want to have a closer look at these equations in order 
to demonstrate that they have the structure of the known 
renormalization group equations in (\ref{RGM}).

The coefficients $c_i(M^2)$ in (\ref{c1P}) and (\ref{c2P}) include only pseudoscalar meson contributions. We will include vector meson contributions soon.
The replacement of the leading $M^2$ dependence by 
\be\label{replacement}
\hat M^2= \ln(2) M^2
\ee
in our previous papers follows from the replacement of $I_{2,3}$ by $\hat I_{2,3}$ in the chiral limit and 
allows us to combine within a very good approximation these results with $z_i$ in the ${\rm \overline{MOM}}$ scheme.
The 
argument of the logarithmic terms is only an approximation since the mass 
scale $\tilde m$ replaces a rather complicated dependence of the exact 
expressions on the meson masses: $m_\pi\le \tilde m \le m_K$. As in our 1986 analysis we set
\be
\tilde m= 0.3\gev~,
\ee
although our results are not very sensitive to this choice unless $\tilde m$ is 
approaching $m_\pi$. In fact it turns out that the matching between quark-gluon and meson evolutions is best for this value. The 
numerical values of $c_{1,2}(M^2)$ for $M=(0.6 - 1.0)\gev$ resulting from 
(\ref{c1P}) and (\ref{c2P}) are given in Table~\ref{tab:Results}.

In (\ref{ME1}) and (\ref{ME2}),  $Q_{1,2}$(0) denote the hadronized $\Delta S =1$ operators at the factorization scale now defined by $\mu=0$.    
As a consequence, the only non-vanishing hadronic matrix elements of current-current operators for the $K\to\pi\pi$ decay amplitudes at $\mu=0$ 
are
\be\label{IC1}
\langle\pi^+\pi^-|Q_2(0)|K^0\rangle=-\langle\pi^0\pi^0|Q_1(0)|K^0\rangle=X_F,
\ee
\be\label{IC2}
\langle\pi^+\pi^0|Q_1(0)|K^+\rangle=\langle\pi^+\pi^0|Q_2(0)|K^+\rangle=\frac{X_F}{\sqrt{2}},
\ee
where
\be\label{XF}
X_F=\sqrt{2}F_\pi(m_K^2-m_\pi^2)
\ee
with the subscript $F$ standing for factorization. Here we keep $m_\pi\not=0$ as the limit $m_\pi\to 0$ is used only for 
operator evolution.
Note that these leading hadronic matrix elements do {\it not} include the usual Fierz terms that are a part of non-factorizable loop corrections. 

The inclusion of the $\ord(1/N)$  non-factorizable loop corrections, represented by the non-vanishing coefficients $c_i$, can 
be viewed as taking into account the physics contributions in  the momentum 
range from $\mu=0$ to $\mu=M$. This is complementary to the  usual 
renormalization group evolution for the Wilson coefficients $z_i$ taking into 
account the physics contributions from $\mu=M$ to $\mu=M_W$. In this manner, 
all physics contributions to the amplitudes ${\rm Re}A_0$ and ${\rm Re}A_2$
from the momentum range from $\mu=0$ to $\mu=M_W$ are 
included.
The inferred pattern for the $Q_{1,2}$ meson evolution has been confirmed by a background field method \cite{Fatelo:1994qh} acting directly at the operator level.

The numerical implications of these results for ${\rm Re}A_0$ and  ${\rm Re}A_2$ 
will be discussed in Section~\ref{sec:5a} but already now we can 
verify that the structure of the equations (\ref{ME1}) and (\ref{ME2})
allows a plausible matching of the meson and quark-gluon evolutions. To this 
end, we have to include in our discussion not only the QCD penguin operator 
$Q_6$ but also $Q_4$ defined in (\ref{Q4}). Its hadronic matrix element at the 
factorization scale is given by
\be\label{IC4}
\langle\pi^+\pi^-|Q_4(0)|K^0\rangle=\langle\pi^0\pi^0|Q_4(0)|K^0\rangle=X_F.
\ee

Then the $4\times4 $ anomalous dimension matrix in the $Q_{1,2,4,6}$ basis, which through 
(\ref{RGM}) governs the evolution of operators in the quark-gluon ($QG$) picture,  reads \cite{Bardeen:1986uz}:
\begin{equation}\label{2.67}
\gamma^{QG} =\frac{\alpha_s N}{2\pi}
\left(\begin{array}{cccc}
0 & 3/N  &  0 & 0\\
3/N & 0  & 1/3N & 1/3N\\
0& 0 & 0 &  0 \\
0& 0 & 0 &  -3
\end{array}\right) = \left(\begin{array}{cccc}
0 & 0.286  &  0 & 0\\
0.286 & 0  & 0.032 & 0.032\\
0& 0 & 0 &  0 \\
0& 0 & 0 &  -0.859
\end{array}\right)
\end{equation}
in the large-$N$ limit. Recall that $\alpha_s N$ is $N$-independent to preserve 
asymptotic freedom in large $N$ QCD. The numerical values above have been 
obtained for $\alpha_s=0.6$, namely around the scale $0.8\gev$ (see Table~\ref{tab:Results2}). This will allow us  a comparison of the 
meson and quark-gluon evolutions.

Using  the evolutions (\ref{ME1}) and (\ref{ME2}) and evaluating  derivatives of $Q_1(M^2)$ and $Q_2(M^2)$ with respect to $M^2$,
we find first
\be\label{M1}
M^2\frac{d\,Q_1(M^2)}{d M^2}=-M^2\frac{d\,c_1(M^2)}{d M^2} Q_2(0),
\ee

\be\label{M2}
M^2\frac{d\,Q_2(M^2)}{d M^2}=-M^2\frac{d\,c_1(M^2)}{d M^2} Q_1(0)+M^2\frac{d\,c_2(M^2)}{d M^2}[Q_2(0)-Q_1(0)].
\ee

But
\be\label{Q60}
Q_4(0)=[Q_2(0)-Q_1(0)], \qquad 
Q_6(0)=- \frac{r^2(\mu)}{\Lambda_\chi^2} [Q_2(0)-Q_1(0)]
\ee
 in our octet approximation. Thus 
\be
Q_4(0)+Q_6(0)=\left(1-\frac{r^2(\mu)}{\Lambda_\chi^2}\right)[Q_2(0)-Q_1(0)].
\ee

Therefore, comparing 
(\ref{M1}) and (\ref{M2}) with (\ref{RGM}) 
 for $\mu=M$, we find the non-vanishing elements of the ''anomalous dimension 
matrix'' $\gamma^{M}$ governing the  evolution of operators in the meson (M) picture:
\be
\gamma^{M}_{12}=\gamma^{M}_{21}= 2M^2\frac{\partial c_1(M^2)}{\partial M^2} > 0,
\ee
\be
\gamma^{M}_{24}=\gamma^{M}_{26}=2M^2\frac{\Lambda_\chi^2}{r^2-\Lambda_\chi^2}\frac{\partial c_2(M^2)}{\partial M^2} >0.
\ee
As $c_{1,2}(M^2)=\ord(1/N)$ the signs and the structure of $1/N$ terms in $\gamma^{M}$ are precisely the same as in (\ref{2.67}), but due to $M^2$ dependence 
of $\gamma^{M}_{ij}$ the evolution of operators is faster in the meson evolution  when the meson theory includes only the pseudoscalar octet. The diagonal 
term $\gamma_{66}$ in (\ref{2.67}) is $\ord(1)$ and originates in the $\mu$-dependence of quark masses. As discussed at the end of this section, in this case there is 
a perfect matching between quark-gluon and meson evolutions in the large $N$ 
limit.

In order to complete the calculation of the anomalous dimension matrix in the 
meson theory we still need the value of $r^2/\Lambda_\chi^2$. This value is 
known in our approach and given in (\ref{useful}). Using this value we find 
at $M=\mu=0.8\gev$
\be\label{ANOMP}
\gamma^{M}_{12}=\gamma^{M}_{21}=0.624, \qquad \gamma^{M}_{24}=\gamma^{M}_{26}=0.051~, \qquad  {\rm (P)}.
\ee

We observe that the hierarchy of the elements of the quark-gluon anomalous dimension matrix in (\ref{2.67}) is also found in the corresponding matrix in 
the meson theory. In particular we find
\be\label{match1}
\frac{\gamma^{M}_{12}}{\gamma^{M}_{26}}=12.2, \qquad \frac{\gamma^{QG}_{12}}{\gamma^{QG}_{26}}=9  \qquad  {\rm (P)}
\ee
which is a satisfactory result considering that we have included only pseudoscalar mesons at this level.

We observe that already the inclusion of pseudoscalar mesons allows a 
reasonable matching between the two anomalous dimensions in question.
On the other hand,  as emphasized in \cite{Bardeen:1986vz}, while the vacuum insertion method gives consistent results for the leading in $N$ contributions, viewed as a meson evolution, it completely misrepresents the next-to-leading effects. Indeed in this case 
the usual $1/N$  Fierz terms give
\be
 \langle Q_1(M^2)\rangle_{\rm VIA}=\langle Q_1(0)\rangle+
\frac{1}{N}\langle Q_2(0)\rangle,
\ee
\be
 \langle Q_2(M^2)\rangle_{\rm VIA}=\langle Q_2(0)\rangle+
\frac{1}{N}\langle Q_1(0)\rangle
\ee
and consequently 
\be 
c_1=-\frac{1}{3}, \qquad c_2=0, \qquad ({\rm in~VIA})
\ee
in total disagreement with the structure of quark-gluon evolution.

 In summary, the structure of meson evolution reviewed above
leads to a very simple physical picture \cite{Bardeen:1986vz}.
 The inclusion of the next-to-leading corrections to hadronic matrix elements can be viewed as the evolution of the operators (\textit{meson evolution}) from zero momentum to $M$.     This short but fast evolution is continued above $M$ as a long but slower evolution of Wilson coefficients (\textit{quark-gluon evolution}) by means of the usual QCD renormalization group equations with respect to $\mu$, with the identification (\ref{scales}).

\boldmath\label{VECTORS}
\subsection{Inclusion of Vector Mesons in $c_1(M^2)$ and $c_2(M^2)$}
\unboldmath
In the same manner, as we did in the case of $\hat B_K$, we can easily include vector meson contributions to the coefficients
$c_{1,2}(M^2)$ and consequently into current-current contributions to the 
amplitudes ${\rm Re} A_2$ and ${\rm Re} A_0$. This is related to the fact 
that in the chiral limit the meson evolutions of the $\Delta S = 2$ and $\Delta I = 3/2$ operators are identical. Keeping pseudoscalar masses in the 
pseudoscalar contributions but calculating the vector contributions in the 
chiral limit
 we simply find:
\be\label{c1V}
c_1(M^2)=c_1^P(M^2)-\frac{1}{4 f_\pi^2}\left[\frac{f_\pi}{F_\pi}\right]\left[\frac{9}{2} \hat I_2(m_V^2)-3 m_V^2\hat I_3(m_V^2)\right],
\ee
where the first term including only pseudoscalar contributions is given in 
(\ref{c1P}). Yet, in evaluating this term we have to use $f_\pi(M^2)$ in 
(\ref{Fpitot}) which includes vector meson contributions. The functions 
$\hat I_i$ are given in (\ref{hatI2}) and (\ref{hatI3}).

We also find
\be\label{c2V}
c_2(M^2)=\frac{1}{2}c_1(M^2)+\frac{9}{8}\frac{m_K^2}{(4\pi f_\pi)^2}\left[\frac{f_\pi}{F_\pi}\right]\ln(1+\frac{M^2}{\tilde m^2})
\ee
with $c_1(M^2)$ given in (\ref{c1V}).

The values for these coefficients with and without the inclusion of vector meson contributions are given in Table~\ref{tab:Results}. Similar to the case 
of $f_\pi(M^2)$, we
 observe significant reduction of the scale dependence of $1-c_1(M^2)$ 
relevant for ${\rm Re} A_2$
relative to the pseudoscalar case  which will have profound implications for 
 our numerical analysis of ${\rm Re} A_2$ in the next section. 

With these results at hand, we can now improve the calculation of the anomalous 
dimension matrix in the meson theory. Setting again $M=\mu=0.8\gev$ we find
\be\label{ANOMPV}
\gamma^{M}_{12}=\gamma^{M}_{21}=0.524, \qquad \gamma^{M}_{24}=\gamma^{M}_{26}=0.060
\qquad  {\rm (P+V)}\ee
and
\be\label{match2}
\frac{\gamma^{M}_{12}}{\gamma^{M}_{26}}=8.7, \qquad \frac{\gamma^{QG}_{12}}{\gamma^{QG}_{26}}=9, \qquad  {\rm (P+V)}
\ee
which is a significant improvement over the result in (\ref{ANOMP}). 

 This
 matching of anomalous dimensions is a remarkable feature of our dual approach
 and  might be traced to the existence of AdS/QCD models which do interpolate between the quark and meson pictures - at least for the amplitudes we are considering.   They usually have extra states at higher mass scales but the pseudoscalar and vector mesons are usually an essential part of the duality.

What remains to be done is to analyze how these results depend on 
$M=\mu$. We show this in Table~\ref{tab:matching}. We draw the following 
conclusions from this table:
\begin{itemize}
\item
In the full range of $M$ considered   $\gamma^{M}_{12}$ is by an order 
of magnitude larger than $\gamma^{M}_{26}$ as is the case in the quark-gluon 
matrix.
\item
If only pseudoscalar mesons are included the ratio 
$\gamma^{M}_{12}/\gamma^{M}_{26}$ is closest to $9$ for $M\approx0.7\gev$, 
while after the inclusion of vector meson contributions this happens 
slightly above  $M\approx0.8\gev$. Therefore we conclude that most 
reliable results are obtained in our approach for $M=(0.8-0.9)\gev$.
\item
Comparing the size of the matrix elements in  Table~\ref{tab:matching} with 
those in (\ref{2.67}) we indeed confirm that the short meson evolution is faster 
than the long quark-gluon evolution. 
\end{itemize}

\begin{table}[!tb]
\centering
\begin{tabular}{|c|c|c|c|c|c|c|}
\hline
 $M[\gev]$  & $ 0.6$ & $0.7$ & $0.8$ & $ 0.9$&  $1.0$ & Comments \\
\hline
\hline
 \parbox[0pt][1.6em][c]{0cm}{}$c_1(M^2)$ & $0.240$  & $0.315$& $0.392$ &$0.471$&$0.549$ & (P)   \\
\parbox[0pt][1.6em][c]{0cm}{}$c_1(M^2)$ & $0.206$  & $0.267$& $0.331$ &$0.398$&$0.468$ & (P+V)\\
\hline
 \parbox[0pt][1.6em][c]{0cm}{}$1-c_1(M^2)$ & $0.760$  & $0.685$ & $0.608$ &$0.529$&$0.451$ & (P)   \\
\parbox[0pt][1.6em][c]{0cm}{}$1-c_1(M^2)$ & $0.794$  & $0.733$ & $0.669$ & $0.602$&$0.532$ & (P+V)\\
\hline
 \parbox[0pt][1.6em][c]{0cm}{}$c_2(M^2)$ & $0.390$  & $0.447$ & $0.498$&$0.543$&$0.584$& (P)   \\
\parbox[0pt][1.6em][c]{0cm}{}$c_2(M^2)$ & $0.390$  & $0.453$ & $0.511$&$0.566$&$0.619$& (P+V)\\
\hline
\end{tabular}
\caption{Values of $c_{1,2}$ as functions 
of $M$. P and V indicate whether pseudoscalar and  vector mesons 
have been included or left out. 
}\label{tab:Results}~\\[-2mm]\hrule
\end{table}

\begin{table}[!tb]
\centering
\begin{tabular}{|c|c|c|c|c|c|c|}
\hline
 $M[\gev]$  & $ 0.6$ & $0.7$ & $0.8$ & $ 0.9$&  $1.0$ & Comments \\
\hline
\hline
 \parbox[0pt][1.6em][c]{0cm}{}$\gamma^{M}_{12}$ & $0.437$  & $0.534$& $0.624$ &$0.706$&$0.784$ & (P)   \\
\parbox[0pt][1.6em][c]{0cm}{}$\gamma^{M}_{26}$ & $0.072$  & $0.059$& $0.051$ &$0.046$&$0.043$ & (P)\\
 \parbox[0pt][1.6em][c]{0cm}{}$\gamma^{M}_{12}/\gamma^{M}_{26}$& $6.0$  & $9.0$ & $12.2$ &$15.3$&$18.3$ & (P)   \\
\hline
 \parbox[0pt][1.6em][c]{0cm}{}$\gamma^{M}_{12}$ & $0.356$  & $0.438$& $0.524$ &$0.615$&$0.714$ & (P+V)   \\
\parbox[0pt][1.6em][c]{0cm}{}$\gamma^{M}_{26}$ & $0.076$  & $0.066$& $0.060$ &$0.058$&$0.057$ & (P+V)\\
 \parbox[0pt][1.6em][c]{0cm}{}$\gamma^{M}_{12}/\gamma^{M}_{26}$& $4.7$  & $6.7$ & $8.7$ &$10.6$&$12.4$ & (P+V)   \\
\hline
\end{tabular}
\caption{Values of $\gamma^{M}_{12}$ and $\gamma^{M}_{26}$  as functions 
of $M$. P and V indicate whether pseudoscalar and  vector mesons 
have been included or left out. 
}\label{tab:matching}~\\[-2mm]\hrule
\end{table}

\boldmath
\subsection{Penguin Operators: $\bsi$ and $\bei$}\label{sec:5}\unboldmath
For the matrix elements of QCD penguin operator $Q_6$ and the electroweak 
penguin operator $Q_8$ we find \cite{Bardeen:1986uz}
\be\label{eq:Q60}
\langle\pi^+\pi^-|Q_6(0)|K^0\rangle=
- \frac{r^2(\mu)}{\Lambda_\chi^2} X_F \,B_6^{(1/2)}=
-\,4 \sqrt{2} 
\left[ \frac{m_{\rm K}^2}{\ms(\mu) + \md(\mu)}\right]^2
\frac{F_\pi}{\kappa} \,B_6^{(1/2)} \, ,
\ee
with the same result for $K^0\to\pi^0\pi^0$ matrix element
and \cite{Buras:1987qa}
\be
\langle\pi^+\pi^0|Q_8(0)|K^+\rangle
=3\left[ \frac{m_{\rm K}^2}{\ms(\mu) + \md(\mu)}\right]^2
F_\pi \,B_8^{(3/2)} \, ,
\label{eq:Q82}
\ee
where 
\be
\kappa=\frac{\Lambda_\chi^2}{m_K^2-m_\pi^2}=\frac{F_\pi}{F_K-F_\pi}=4.93.
\ee
In (\ref{eq:Q60}) and (\ref{eq:Q82}) we have introduced the parameters 
$\bsi$ and $\bei$ in order to compare with lattice results. But 
in the large $N$ limit in which factorization works we simply have
 as seen from   (\ref{IC1}), (\ref{IC2}) and (\ref{Q60})
\be\label{bsibei}
\bsi=\bei= 1~.
\ee

Finally, for our numerical studies we quote at $\mu=0.8\gev$
\be\label{useful}
\frac{r^2(\mu)}{\Lambda_\chi^2}=8.46 \left(\frac{160\mev}{m_s(\mu)+m_d(\mu)}\right)^2,
\ee
where we used the results from FLAG 2013  \cite{Aoki:2013ldr}
\be
m_s(2\gev)=(93.8\pm2.4) \mev, \qquad
m_d(2\gev)=(4.68\pm0.16)\mev.
\ee

There is no contribution of $Q_6$ to $K^+\to\pi^+\pi^0$ in the isospin limit, 
but in the case of $\epe$ isospin breaking corrections leading to a
non-vanishing matrix element $\langle Q_6\rangle_2$ have to be taken into account  {as implemented in \cite{Buras:1987wc} where the full $0^{-+}$ nonet has been consistently included  at 
$\ord(p^2)$. The most recent  discussion of this issue with
$0^{-+}$ octet at  $\ord(p^4)$ can be found in 
\cite{Cirigliano:2003nn}.}

It should be stressed that generally the parameters $\bsi$ and $\bei$ are 
very weakly dependent on the scale $\mu$ as the dominant $\mu$ dependence of 
the matrix elements of penguin operators
comes from the running 
quark masses. This dependence in physical amplitudes is canceled by the 
$\mu$ dependence of the corresponding Wilson coefficients, which for large 
$N$ can be demonstrated analytically. This cancellation results from the fact 
that the anomalous dimensions of these operators equal twice the anomalous 
dimension of the mass operator. The effect of mixing with other operators (see 
$\gamma_{26}\not=0$)
spoils this exact cancellation but the effect is  small and is compensated 
by contributions from current-current operator $Q_2$. A detailed numerical 
analysis in  \cite{Buras:1993dy} confirms this.

The $1/N$ corrections to the result in (\ref{bsibei}) are not necessary 
for the analysis of the $\Delta I=1/2$ rule, as the Wilson coefficients of 
QCD penguin $Q_6$ are $\ord(\alpha_s)$ and therefore 
these contributions are $\ord(1/N^2)$. 
In the case of $Q_8$ such corrections could possibly play a role in $\epe$.

There is no reliable result on $\bsi$ from lattice QCD. On the other 
hand  one can  extract the lattice value for $\bei$ from ${\rm Re}A_2$ in
\cite{Blum:2012uk}.
We find 
\be
\bei(3\gev)= 0.65\pm 0.05 \qquad {\rm (lattice)}.
\ee

Even if  $\bei$  is scale independent in the large $N$ limit, 
it is useful  to check its scale dependence in the short-distance regime 
by means of renormalization group evolution, this time for the matrix element 
of $Q_8$, not for its Wilson coefficient. Such an exercise has been performed 
in  \cite{Buras:1993dy} with the result that this dependence is at the level 
of a few percent for $1.0\gev\le \mu\le 3\gev$ and even if  $\bei$ decreases with increasing scale the difference
between lattice result and large $N$ result cannot be explained by such effects.
On the other hand, the calculation of $1/N$ corrections to (\ref{bsibei}) 
in the 
framework of truncated meson theory of Section~\ref{sec:2} shows that while 
these corrections are small in the case of $\bsi$, the effect is much larger 
in the case of $\bei$ \cite{Hambye:1998sma}. Typically $\bei$ is found in the ballpark of $0.6\pm 0.1$. Consequently also in this case the large $N$ approach seems to give a result 
similar to the lattice one. Yet, one has to admit that the precision of 
the calculation in \cite{Hambye:1998sma} is insufficient for a useful phenomenology of $\epe$, where there is a strong cancellation between QCD penguin and electroweak penguin contributions.
On the other hand, both lattice calculations and large $N$ approach indicate 
that $\bei< 1$ suppressing electroweak penguin contributions to $\epe$ 
relative to strict large $N$ limit. This is
 a hint that $\epe$ in the Standard Model is larger than previously 
expected. Yet, the future of $\epe$ depends on the result for $\bsi$ in 
lattice QCD although it may still  
take some time before an accurate 
result 
for this important quantity is available \cite{Christ:2013lxa}.

\boldmath
\section{$\Delta I=1/2$ Rule in the Dual QCD Approach} 
\label{sec:5a}
\unboldmath
\subsection{Preliminaries}
With all these results at hand, we will now make a closer look at the 
dynamics of the $\Delta I=1/2$ rule which follows from our approach.
Even without entering the details of the size of the amplitudes involved, 
we note that the amplitude $A(K^0\rightarrow\pi^0\pi^0)$ 
enters ${\rm Re}A_0$ and ${\rm Re}A_2$ in (\ref{DI1}) and (\ref{DI2})
with the opposite sign. While this feature is
at the basis of the difference between $A_0$ and $A_2$ and consequently fundamental for the explanation of the $\Delta I=1/2$ rule, 
the main dynamics behind the $\Delta I=1/2$ rule is that
 $A(K^0\rightarrow\pi^0\pi^0)$ has the 
same sign as $A(K^0\rightarrow\pi^+\pi^-)$. That this is indeed the case 
follows both from the explicit evaluation of 
these amplitudes in our dual representation of QCD and also from 
our simple picture 
of the slow quark-gluon evolution from $\ord(M_W)$ down to $\ord(1\gev)$ 
followed by the fast meson evolution down to $\mu=\ord(m_\pi)$. We will 
now discuss these issues in explicit terms by updating
 our analysis of the amplitudes  ${\rm Re}A_0$ and  ${\rm Re}A_2$ 
presented first in \cite{Bardeen:1986vz}.

\subsection{Hadronic Matrix Elements}
The values for hadronic matrix elements of current-current operators in (\ref{IC1}) and (\ref{IC2}) are simply the initial conditions for the meson evolution, 
analogous to the initial conditions for Wilson coefficients $z_1(M_W^2)$ and 
$z_2(M_W^2)$ in the case of quark-gluon evolution that are usually evaluated 
perturbatively at the high energy scale. Combining then the initial 
conditions in  (\ref{IC1}) and (\ref{IC2})  with the meson evolution 
formulae in (\ref{ME1}) and (\ref{ME2}) allows us to evaluate 
hadronic matrix elements of current-current operators at $\mu=M$
\cite{Bardeen:1986vz}:
\be\label{X1}
\langle\pi^+\pi^-|Q_1(M^2)|K^0\rangle=-c_1(M^2) X_F, \quad
\langle\pi^+\pi^-|Q_2(M^2)|K^0\rangle=\left(1+c_2(M^2)\right)X_F,
\ee
\be\label{X3}
\langle\pi^0\pi^0|Q_1(M^2)|K^0\rangle=-X_F,\quad 
\langle\pi^0\pi^0|Q_2(M^2)|K^0\rangle=\left(c_1(M^2)+c_2(M^2)\right)X_F,
\ee
\be\label{X5}
\langle\pi^+\pi^0|Q_{1,2}(M^2)|K^+\rangle=\left(1-c_1(M^2)\right)\frac{X_F}{\sqrt{2}}.
\ee
where $X_F$ has been defined in (\ref{XF}).

Using (\ref{basic}), (\ref{DI1}), (\ref{DI2}), the matrix elements (\ref{X1})--(\ref{X5})  and (\ref{eq:Q60}) with $\bsi= 1$ we find 
then
\be\label{BBG1}
{\rm Re}A_0=\frac{G_F}{\sqrt{2}}V_{ud}V_{us}^*\left(\frac{1}{3}\right)
\left[-z_1(1+2 c_1)+z_2(2+c_1+3 c_2) - 3 z_6 \frac{r^2}{\Lambda_\chi^2}\right]X_F
\ee

\be\label{BBG2}
{\rm Re}A_2=\frac{G_F}{\sqrt{2}}V_{ud}V_{us}^*\left(\frac{2}{3}\right)
(1-c_1)(z_1+z_2)\frac{X_F}{\sqrt{2}}=2.54~(1-c_1)(z_1+z_2)~10^{-8}\gev,
\ee
where in order to simplify the notations we did not show the scale dependence 
of $z_i$ and $c_k$, explicitly. They all are evaluated at $\mu=M$.

\subsection{Diagrammatic Understanding of Signs}
In order to get a better understanding of different signs in (\ref{BBG1}) and 
(\ref{BBG2})  and eventually 
to compare with the results of the RBC-UKQCD collaboration \cite{Boyle:2012ys,Blum:2011pu,Blum:2011ng,Blum:2012uk}, we will use the diagrammatic language developed by us in the context 
of our first paper on large $N$ approach to weak decays that we applied for
the decays $D^0\to K^+\pi^-$ and $D^0\to \bar K^0 \pi^0$ \cite{Buras:1985xv} \footnote{See Figs. 10-11 in \cite{Buras:1985xv}. Note that the indices of $Q_1$  
and $Q_2$ are interchanged in that paper.}.
This diagrammatic language inspired  by the work of
't Hooft \cite{'tHooft:1973jz,'tHooft:1974hx} and subsequently Witten \cite{Witten:1979kh,Treiman:1986ep} is discussed in detail for the case of non-leptonic $K$
decays in \cite{Buras:1988kp,Buras:1988ky,Gerard:1990dx}. See also 
\cite{Buras:1985yx}. 

It is clear from these papers, but  should be emphasized again, 
that the diagrams discussed by us should not be considered
as ordinary Feynman diagrams as each of the closed loops stands for sum over 
all possible planar gluon exchanges.

\begin{figure}[thb]
\centerline{\includegraphics[width=0.65\textwidth]{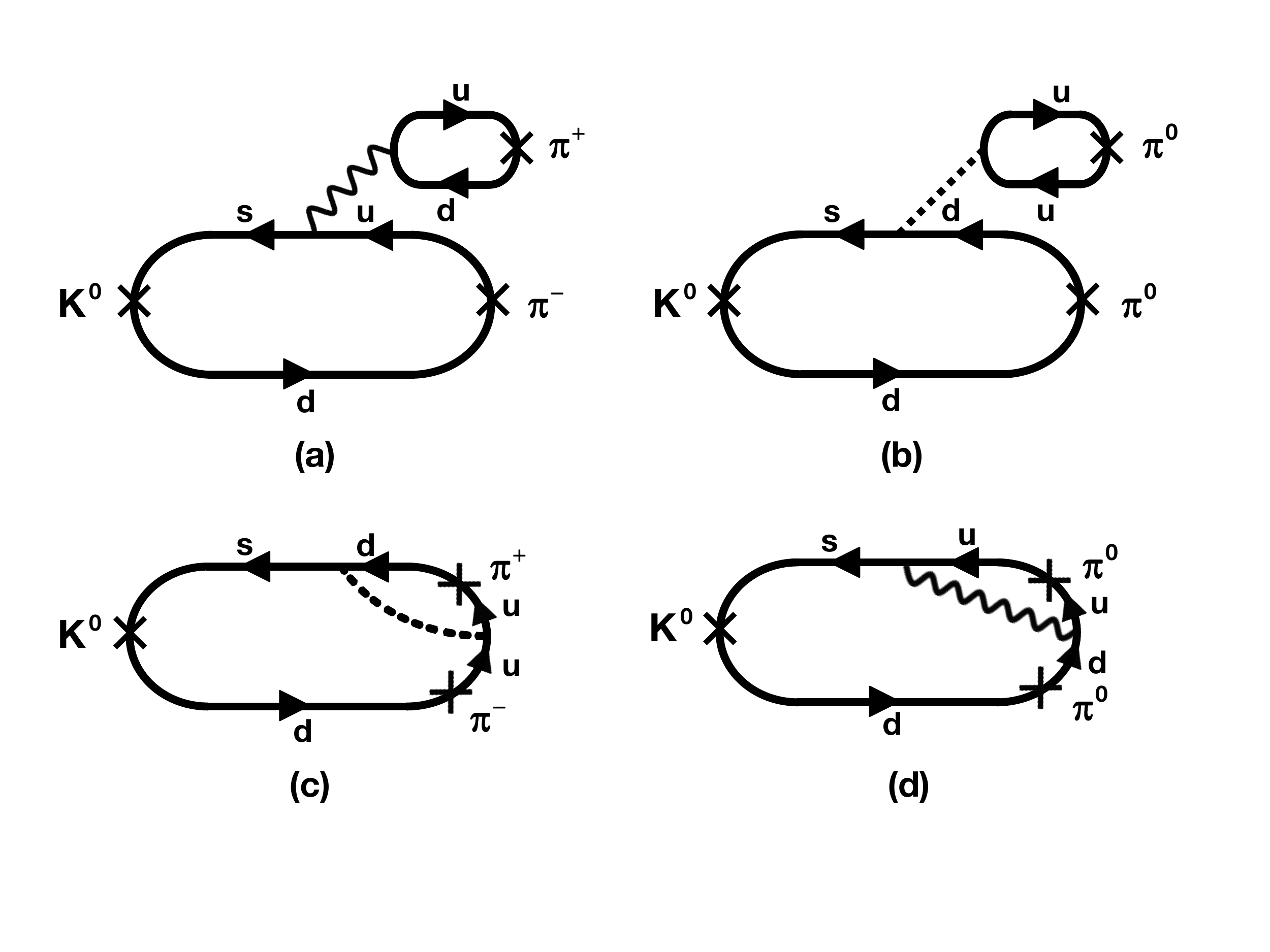}}
\vskip-0.8cm
\caption{Current-current topologies in $K\to\pi\pi$.}\label{Fig:1}
\end{figure}

\begin{figure}[thb]
\centerline{\includegraphics[width=0.65\textwidth]{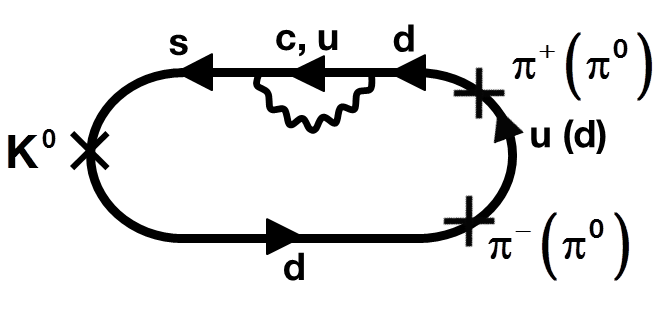}}
\vskip-0.2cm
\caption{Penguin topology in $K\to\pi\pi$.}\label{Fig:2}
\end{figure}

In Fig.~\ref{Fig:1} we show four basic current-current diagrams contributing to 
$K^0\to\pi^+\pi^-$  and $K^0\to \pi^0\pi^0$ amplitudes. The four diagrams
contributing to $K^+\to\pi^+\pi^0$ can be obtained from these 
diagrams by replacing the spectator quark $d$ by the spectator quark $u$.
The wiggly line represents the insertion of the $Q_2$ operator, while the dashed one the insertion of $Q_1$ 
operator. The crosses represent the external mesons. The Feynman rule for 
them is the usual colour normalization factor $1/\sqrt{N}$ but this universal
 rule does not interest us here. It is more important that each loop brings a factor $N$ so that the diagrams (c) and (d) 
are suppressed relative to (a) and (b) by a factor of $N$. In Fig.~\ref{Fig:2}  we 
show penguin diagrams that contribute only to $K^0\to\pi^+\pi^-$  and $K^0\to \pi^0\pi^0$ amplitudes in the isospin limit.

The results for the
matrix elements in (\ref{X1})-(\ref{X5}) can be reproduced from these diagrams 
by using the following Feynman-like rules:

{\bf Rule 1:} Factor $X_F$ for diagrams  (a) and (b) in Fig.~\ref{Fig:1}

{\bf Rule 2:} Factor $c_1X_F$ for diagrams  (c) and (d) in Fig.~\ref{Fig:1}

{\bf Rule 3:} Factor $c_2X_F$ for penguin diagrams in Fig.~\ref{Fig:2}

{\bf Rule 4:} Statistical factor $-1$ for each quark loop.

{\bf Rule 5:} Factor $-1$  when the final neutral pion is created
                through its $\bar d d$ component.

{\bf Rule 6:} Factor $-1$ in the penguin diagrams due to the GIM 
              partial cancellation at work ($V_{cd}V_{cs}^*=-V_{ud}V_{us}^*$).

As the factorizable contribution $X_F$ is positive, the dynamics of the $\Delta I=1/2$ rule is governed 
by the non-factorizable topologies in (c) and (d) represented by the 
coefficient $c_1(M^2)$ and it is essential that this coefficient is also
positive. In our approach, this positive sign follows in two ways:
\begin{itemize}
\item
From explicit calculation of loop diagrams in the meson theory.
\item
From the matching of anomalous dimensions $\gamma^{QG}$ and $\gamma^{M}$.
\end{itemize}

This understanding of the sign of $c_1(M^2)$ will allow us in Section~\ref{Lattice} to understand the signs of contractions in the recent results on
${\rm Re}A_0$  on ${\rm Re}A_2$ 
 from the RBC-UKQCD collaboration \cite{Boyle:2012ys}. But first we present 
our own view on these amplitudes.

\boldmath
\subsection{The Anatomy of the $\Delta I=1/2$ Rule}
\unboldmath
We are now ready to have a closer look at the basic dynamics behind this rule 
which in our approach is based on two pillars of QCD: asymptotic freedom 
at short distance scales and confinement of quarks in mesons at long 
distance scales. The dual representation of QCD as a theory of weakly interacting mesons allows to unite these two properties in a framework which enables 
us to see and calculate analytically the observed enhancement of ${\rm Re}A_0$  and suppression of ${\rm Re}A_2$.
\begin{itemize}
\item
In the strict large $N$ limit, no evolution takes place:
\be
z_1=0, \quad z_2=1.0, \quad z_6=0, \quad c_1=c_2=0.
\ee
Then only the operator $Q_2$ contributes and
its factorized hadronic 
matrix elements imply a vanishing $K^0\to\pi^0\pi^0$ decay amplitude. Consequently 
\be\label{R02}
R=\frac{{\rm Re}A_0}{{\rm Re}A_2}=\sqrt{2}, \qquad ({\rm in~large}~ N~{\rm  limit})
\ee
in plain disagreement with experiment. The same applies for separate amplitudes 
as seen in (\ref{LO}). 

 In this starting point hadronic matrix 
elements are evaluated in the free meson theory which 
corresponds to the factorization scale $\mu=\ord(m_\pi)$, while the Wilson 
coefficients are calculated in a free (from the point of view of strong interactions) theory of quarks which corresponds to scales $\mu=\ord(M_W)$ and 
setting $\alpha_s(M_W)=0$. In the following steps the gap between these two 
vastly different energy scales is filled with the QCD dynamics present in quark-gluon 
and meson evolutions.
\item
The inclusion of a long but slow logarithmic quark-gluon evolution from 
$\mu=M_W$ down to $\mu=\ord(1\gev)$, termed in the past as {\it octet enhancement}
\cite{Gaillard:1974nj,Altarelli:1974exa}, generates the operator $Q_1$ and 
modifies $z_2$ so that now
\be
z_1<0, \quad z_2>1.0, \quad z_6=0, \quad  c_1=c_2=0,
\ee
where we did not include yet QCD penguin contribution.

Evaluating the Wilson coefficients of $Q_1$ and $Q_2$  
at a scale $\ord(1\gev)$ but keeping their hadronic matrix elements at $\mu=0$, 
we find an enhancement of $R$ by roughly a factor of two in the NDR-${\rm \overline{MS}}$ scheme,  but more like three in the ${\rm \overline{MOM}}$ scheme. This difference is then canceled 
by the scheme dependence of hadronic matrix elements but this fact shows 
that the size of this enhancement attributed to quark-gluon evolution (Wilson coefficients) and separately to meson evolution (hadronic matrix elements) is both dependent on $\mu$ and the renormalization scheme considered. As we use ${\rm \overline{MOM}}$ scheme in our paper, we quote using Table~\ref{tab:Results2}
\be
R_{\rm cc}(\mu)=\sqrt{2}\left(\frac{z_2(\mu)-z_1(\mu)/2}{z_2(\mu)+z_1(\mu)}\right)\approx 3.1\sqrt{2}\approx 4.4 \qquad (\mu =0.8 \gev),
\ee
where with the index ``cc'' we indicate that only current-current operator 
contributions have been taken into account. In the NDR-${\rm \overline{MS}}$ 
scheme we find $R_{\rm cc}\approx 3$ instead. For the amplitudes at this stage 
we find in the ${\rm \overline{MOM}}$ scheme at $\mu=0.8\gev$
\be\label{Step1}
{\rm Re}A_0=7.1\times 10^{-8}\gev ,\qquad   {\rm Re}A_2= 1.6\times 10^{-8}\gev~, 
\qquad ({\rm QG~ evolution}).
\ee
This means an enhancement of ${\rm Re}A_0$ by a factor of $2.0$ and suppression of  ${\rm Re}A_2$ by a factor of $1.6$ relative to the large $N$ limit 
values in (\ref{LO}). While this result is very encouraging, we should note 
that out of the missing factor of 15.8 for $R$ in the large $N$ limit we 
have explained only $3.2$. Therefore we have to include also QCD dynamics below
 $\mu=0.8\gev$.
\item
In our approach, 
switching next the short but fast quadratic meson evolution from $\mu=0$ to $\mu=\ord(1\gev)$ in order to match the quark evolution provides additional enhancement of 
  ${\rm Re}A_0$ and additional suppression of  ${\rm Re}A_2$ due to 
positive values of the coefficients $c_1$ and $c_2$:
\be\label{Rcc}
R_{\rm cc}=\sqrt{2}\left\{\frac{z_2(1+c_1/2+3 c_2/2)-(z_1/2)(1+2 c_1)}{(z_2+z_1)(1-c_1)}
  \right\} \approx 12.4,
\ee
where we quoted the value obtained for $\mu=0.8\gev$. This is only 
$40\%$ below the experimental value in (\ref{N1a}) but does not yet include 
penguin contributions that will enhance $R$ in the direction of experimental 
value. Yet, already with this dynamics we succeded to explain the factor $8.7$ 
out of required factor of $15.8$, that is an order of magnitude enhancement of 
$R_{\rm cc}$ of which $3.1$ is attributed to QG evolution and $2.8$ to the 
 M evolution.

 For the amplitudes, at this stage 
we find in the ${\rm \overline{MOM}}$ scheme at $\mu=0.8\gev$
\be\label{Step2}
{\rm Re}A_0=13.3\times 10^{-8}\gev ,\qquad   {\rm Re}A_2= 1.07\times 10^{-8}\gev~, 
\qquad ({\rm QG+M ~ evolution})
\ee

We would also like to emphasize that
for $c_2=0$  the amplitude 
${\rm Re}A_2$ would remain unchanged but  ${\rm Re}A_0$ 
 would decrease relative to (\ref{Step2})
\be\label{Step3}
{\rm Re}A_0=9.1\times 10^{-8}\gev ,\qquad   R_{cc}=8.5, 
\qquad ({\rm QG+M ~ evolution,~c_2=0}).
\ee
This tells us that the presence of mixing between $Q_2$ and $Q_6$ operators represented by $c_2$ in the meson theory plays a larger role than  $c_1$ in enhancing  ${\rm Re}A_0$ but has no impact on  ${\rm Re}A_2$.
\item
Finally the contribution from the penguin operators, 
in particular from $Q_6$, pointed out in \cite{Shifman:1975tn}, has to be taken into 
account. This operator 
contributes only to ${\rm Re}A_0$ in the isospin limit. Its Wilson coefficient $z_6$ is negative and GIM suppressed for $\mu$ significantly larger than $m_c$. 
But as shown in \cite{Shifman:1975tn} if it is evaluated at $\mu$ as low as few hundreds $\mev$, it then provides 
the dominant bulk of the enhancement of ${\rm Re}A_0$. Even if perturbation theory breaks down at such low scales, it is evident from our approach that at 
such scales the meson evolution has only very minor contribution to matrix 
elements of current-current operators. In our case we have for $\mu=\ord(1\gev)$
\be\label{Rp}
R_p(\mu)=\frac{3\sqrt{2}}{2}\left(-\frac{r^2(\mu)}{\Lambda^2_\chi}\right)\left(\frac{z_6(\mu)}{z_2(\mu)+z_1(\mu)}\right).
\ee
At these scales the QCD penguin contribution to  ${\rm Re}A_0$ gets smaller 
than in  \cite{Shifman:1975tn}, but as we will see below it is still 
significant.
\end{itemize}

In summary taking all effects into account we end up with 
\be
R_{\rm tot}=R_{\rm cc}+\frac{R_p}{(1-c_1)}~,
\ee
where $R_p$ and $R_{\rm cc}$ are given in (\ref{Rp}) and (\ref{Rcc}), 
respectively.
 We  emphasize again that the relative size of 
current-current and QCD penguin contributions to  ${\rm Re}A_0$  depends on the   matching    scale $\mu$ 
considered, and the QCD penguin contribution decreases with increasing $\mu$. 
While in our case the latter contribution will amount to more than $15\%$ 
of  ${\rm Re}A_0$, in lattice calculations that work at 
$\mu=(2-3)\gev$, current-current contributions dominate by far and the trace of a significant QCD penguin contributions found in our case at lower $\mu$ 
should be found in the hadronic matrix element of the  current-current operator $Q_2$. Clearly 
the final amplitudes cannot depend on the chosen matching scale, but relative contributions are $\mu$-dependent. 

With this insight, before presenting graphically the budget of various contributions in Fig.~\ref{fig:pies}, 
we will present the results for ${\rm Re}A_0$ and  ${\rm Re}A_2$  for 
other values of $M$ with and without  vector meson contributions but 
always  matching in the ${\rm \overline{MOM}}$ schemes as well as using the input of 2014.

\subsection{Numerical Analysis.}
In Table~\ref{tab:input} we give the values of various quantities that 
we kept fixed in our analysis. In particular the value of  $m_s$
relevant for QCD penguin contribution has been evaluated at $\mu=0.8\gev$. 
The values of $c_{1,2}(M^2)$ including and leaving out vector meson 
contributions are given in Table~\ref{tab:Results}. 

Before presenting our results we would like to address the following problems 
and state our solutions to them: 
\begin{itemize}
\item
Concerning meson evolution, in the case when only pseudoscalars are included, 
our results can only be trusted up to the scale $M=0.6\gev$. When 
vector mesons are included this range can be extended to scales 
$M=(0.8 - 0.9)\gev$. 
\item
Concerning quark-gluon evolution one would ideally stop it around the scales 
explored by lattice calculations, that is $\mu=(2-3)\gev$. But this is clearly
impossible in our approach and we have to evaluate the coefficients at 
scales $\mu$ as low as $1\gev$ and even $0.8\gev$. As explained above we 
have evaluated the Wilson coefficients at NLO in the ${\rm \overline{MOM}}$ scheme, which is 
the scheme to be used to match with the meson evolution. 
The differences between ${\rm \overline{MOM}}$ scheme 
and NDR-${\rm \overline{MS}}$  as shown in Table~\ref{tab:Results2} are 
 sizable with the short distance effects being significantly larger in the 
${\rm \overline{MOM}}$ scheme. Therefore in this scheme, as demonstrated 
already, 
the $\Delta I=1/2$ rule is more 
visible in the Wilson coefficients than in the NDR-${\rm \overline{MS}}$
scheme used by lattice groups. This difference must then be compensated by 
the corresponding values of hadronic matrix elements.
\end{itemize}

When vector meson contributions are taken into account, but higher resonances 
are not included, it is plausible that 
the optimal matching scale is  $M=\mu=0.8$. Indeed at this scale  the evaluation of both 
the  contributions from meson and quark-gluon evolutions can be trusted, even 
if we cannot claim precision. 
 However, it will be instructive to provide the results  also for the full range
of $0.6\gev\le\mu=M\le 1.0\gev$  with and without  
vector meson contributions in order to see how good the matching is.

\begin{table}[!tb]
\centering
\begin{tabular}{|c|c|c|c|c|c|c|c|}
\hline
 $M=\mu[\gev]$  & $ 0.6$ & $0.7$ & $0.8$ & $ 0.9$&  $1.0$ & Comments& Data \\
\hline
\hline
\parbox[0pt][1.6em][c]{0cm}{}$10^{8}{\rm Re} A_2[\gev]$ &$1.06$  &$1.04$ & $0.97$&$0.88$& $0.77$&  (P) & $1.21$ \\
\parbox[0pt][1.6em][c]{0cm}{}$10^{8}{\rm Re} A_2[\gev]$ &$1.11$  &$1.11$ & $1.07$&$1.00$ & $0.91$ &  (P+V)  &$1.21$      \\
\parbox[0pt][1.6em][c]{0cm}{}$10^{8}{\rm Re} A_0[\gev]$ ($cc$) & $14.2$& $13.7$  &$13.6$& $13.6$ &$13.7$ & (P)& $27.0$ \\
\parbox[0pt][1.6em][c]{0cm}{}$10^{8}{\rm Re} A_0[\gev]$ ($cc$) & $13.9$& $13.4$  &$13.3$& $13.4$ &$13.6$ & (P+V) & $27.0$ \\
\parbox[0pt][1.6em][c]{0cm}{}$R_{cc}$& $13.4$ &$13.2$& $14.0$ &$15.5$&$17.8$& (P) & $22.4$    \\
\parbox[0pt][1.6em][c]{0cm}{}$R_{cc}$& $12.5$ &$12.0$& $12.4$ &$13.4$&$14.9$& (P+V) & $22.4$     \\
\hline
\end{tabular}
\caption{The anatomy of the current-current contributions to the $\Delta I=1/2$ rule as function of the matching scale. P and V indicate that pseudoscalar and  vector mesons 
have been included. 
}\label{tab:DIRule}~\\[-2mm]\hrule
\end{table}

In Table~\ref{tab:DIRule} we show the results for ${\rm Re} A_2$ and
 ${\rm Re} A_0$ including only current-current contributions and calculating 
Wilson coefficients in the ${\rm \overline{MOM}}$ scheme.
We indicate by P and V which  meson contributions have been included. 
We observe:
\begin{itemize}
\item
For scales $M\approx 0.8\gev$  ${\rm Re} A_2$ is typically 
suppressed by a factor of $2.4$ relative to the strict large $N$ limit, which is
slightly more than required by the data. Moreover, in the absence of vector meson contributions 
${\rm Re} A_2$ drops quickly down with increasing $M$. The inclusion of 
vector meson contributions softens significantly this suppression. Even 
if at $\mu=0.8\gev$ the amplitude ${\rm Re} A_2$ is found
 by $12\%$ below the experimental value, this result should be considered  
as a success of our approach. Indeed ${\rm Re} A_2$  is  rather  close to the data after the vector meson contributions have been included. This allows us to expect that a more complete treatment including heavier resonances could further improve the matching conditions and agreement with experiment.
\item
The amplitude ${\rm Re} A_0$ turns out to be rather insensitive to the 
inclusion of vector contributions. At $M=0.8\gev$ roughly $50\%$ of its 
experimental value is described by current-current contributions. This 
could appear disappointing but one should remember that in the case of the
$K^0\to\pi^0\pi^0$ amplitude only the first non-vanishing term in $1/N$ 
expansion has been included. Still ${\rm Re} A_0$ is enhanced by a factor 
of $3.7$ over its leading term which should be regarded as a significant achievement. Moreover, as we will see soon, QCD penguin contributions help bring  ${\rm Re} A_0$ closer to the data.
\end{itemize}

\begin{table}[!tb]
\centering
\begin{tabular}{|c||c|c|c|c|c|c|c|c|}
\hline
  $M [\gev]$ & $0.6$ &  $0.7$ &$0.8$ & $0.9$   &   $1.0$ & Scheme \\
\hline
$c_1$ & $0.133$ & $0.201$ &  $0.244$   &   $0.272$ & $0.295$ & ${\rm \overline{MOM}}$ \\
$c_1$ & $0.338$ & $0.355$ &  $0.369$   &   $0.379$ & $0.389$ &  NDR-${\rm \overline{MS}}$ scheme  \\
\hline
\end{tabular}
\caption{The values of $c_1 (M^2)$ extracted from the  data on      ${\rm Re}A_2$ for different values of the matching scale in   ${\rm \overline{MOM}}$  and  NDR-${\rm \overline{MS}}$ scheme .
}\label{tab:C1EXP}~\\[-2mm]\hrule
\end{table}

Concerning ${\rm Re} A_2$ we may ask what are the values of $c_1(M^2)$ that would reproduce 
exactly the experimental value of ${\rm Re} A_2$. Following (\ref{BBG2}), such values are given by 
\be
c_1(M^2)=1-\frac{0.476}{(z_1(M)+z_2(M))}.
\ee
We show the result of this exercise in Table~\ref{tab:C1EXP}. We emphasize 
the scheme dependence of this result.

\begin{table}[!tb]
\centering
\begin{tabular}{|c||c|c|c|c|c|c|c|c|}
\hline
 \parbox[0pt][1.6em][c]{0cm}{} $|z_6|\bsi$ &  $0.04$ &$0.06$ & $0.08$ & $0.10$ & $0.12$ & $0.14$ & $0.20$ & Data \\
\hline
\parbox[0pt][1.6em][c]{0cm}{}$10^{8}{\rm Re} A_2[\gev]$ & $1.07$ &$1.07$ &$1.07$ &$1.07$  &$1.07$ &$1.07$ & $1.07$ &  $1.21$ \\
\parbox[0pt][1.6em][c]{0cm}{}$10^{8}{\rm Re} A_0[\gev]$ (tot) &$15.1$  &$16.0$  & $17.0$ & $17.9$ & $18.8$ & $19.7$  & $22.4$ &  $27.0$ \\
\parbox[0pt][1.6em][c]{0cm}{}$R_{tot}$& $14.1$  & $15.0$  & $15.8$ & $16.7$  & 
$17.5$ & $18.4$ & $20.9$ & $22.4$ \\
\hline
\end{tabular}
\caption{${\rm Re} A_2$, ${\rm Re} A_0$  and $R_{tot}$ including QCD penguin contribution for different values of $|z_6|\bsi$ and the matching scale $M=0.8\gev$.  Both $P$ and $V$ are included.  ${\rm \overline{MOM}}$ scheme 
for $z_i$ has been used.
}\label{tab:final}~\\[-2mm]\hrule
\end{table}

In order to complete the analysis we have to include QCD penguin contributions 
which further enhance ${\rm Re} A_0$. In this context, we would like to 
recall the analysis  in \cite{Bardeen:1986uz} where the effects of 
an incomplete GIM mechanism above $m_c$ on the mixing between current-current 
operators and the value of $z_6$ have been estimated. It has been found that 
these effects could, at scale $\mu=(0.8 -1.0)\gev$, enhance $|z_6|$ by a factor  
of $2-3$ relative to the leading order result in which GIM is assumed to be exact for 
$\mu\ge m_c$. In Table~\ref{tab:final} we show our 
final result at $M=\mu=0.8\gev$, including vector meson contributions for different values of the product $|z_6|\bsi$.  Its value $0.04$ corresponds to exact GIM mechanism above $m_c$  and 
$\bsi=1.0$. The remaining values $0.06-0.14$ correspond to the effect of 
incomplete GIM mechanism above $m_c$ estimated in \cite{Bardeen:1986uz} and/or 
  values  $\bsi$ above unity. However as seen in 
 Table~\ref{tab:final} even for $|z_6|\bsi=0.20$ the experimental 
 value of ${\rm Re} A_0$ cannot be fully reproduced.

In summary, we observe that our approach  provides an order of magnitude 
enhancement of $R$ relative to the strict large $N$ result $R=\sqrt{2}$. 
This enhancement follows mainly from the suppression of ${\rm Re} A_2$ by a factor of
$2.4$ and the enhancement of ${\rm Re} A_0$ by a factor of $3.7$ from 
current-current contributions. In this manner we improve significantly 
on the original work on octet enhancement \cite{Gaillard:1974nj,Altarelli:1974exa} where only quark-gluon evolution has been taken into account and the 
result was scale and renormalization scheme dependent.
Including QCD penguin contributions the 
latter enhancement increases to $4$ in the case of exact GIM above $\mu=m_c$ 
and could be even as high as $5$ if the effects of incomplete GIM are 
taken into account. In this manner the main bulk (factor $10-12$) of the observed enhancement of $R$ relative to $R=\sqrt{2}$ by a factor of $15.8$ can be explained.

On 
the other hand  while ${\rm Re} A_2$  is found only $12\%$ below the data,  ${\rm Re} A_0$ is  found to  be $40\%$ smaller than its measured value when strict GIM mechanism is assumed above the charm scale. 
Our analysis shows therefore that at the scales we are working 
QCD penguin dynamics  in the amplitude ${\rm Re} A_0$  is relevant. The missing $40\%$ in ${\rm Re} A_0$ can be attributed to the effects of higher resonances in the meson evolution for current-current operators, higher $1/N$ corrections and, as stated above 
and seen in  Table~\ref{tab:final},  to non-GIM effects above the charm scale and increased value of $\bsi$. As our calculation of vector meson contributions 
has been performed in the chiral limit also here some improvements are possible.
A full AdS/QCD description should be able to provide a more complete picture of the long distance terms and the matching of the amplitudes to the expected short distance behaviour.  It would also constrain any purely non-perturbative contribution not directly accessible through current-current operator evolution and matching, such as the one from the $\Delta I=1/2$ weak mass operator 
$(mU^\dagger+{\rm h.c.})_{ds}$  coupled to either the gluonic term \cite{Gerard:2000jj} or the quark mass term of the strong trace anomaly \cite{Crewther:2013vea}.

For higher matching scales as $\mu=(2-3)\gev$, 
used in lattice calculations, the role of QCD penguins in ${\rm Re} A_0$ 
will be much smaller. The incomplete GIM effects above $m_c$ discussed here 
should then be found dominantly in the enhanced hadronic matrix elements of 
current-current operators, in particular $Q_2$. The comparison with latest lattice analyses is given in the next section.

\begin{figure}[t]
\begin{center}
\includegraphics[height=6.0cm]{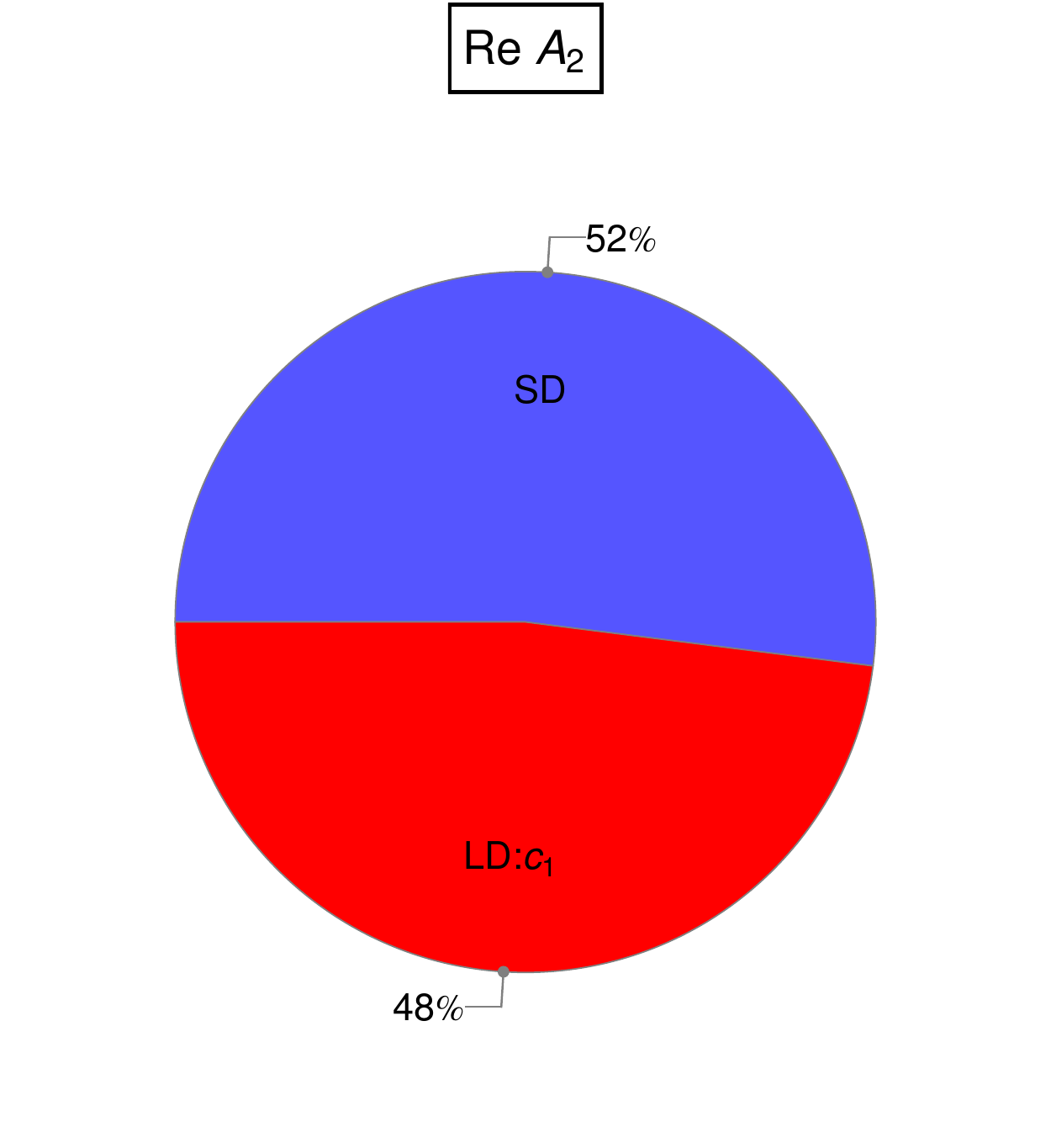}
\includegraphics[height=6.0cm]{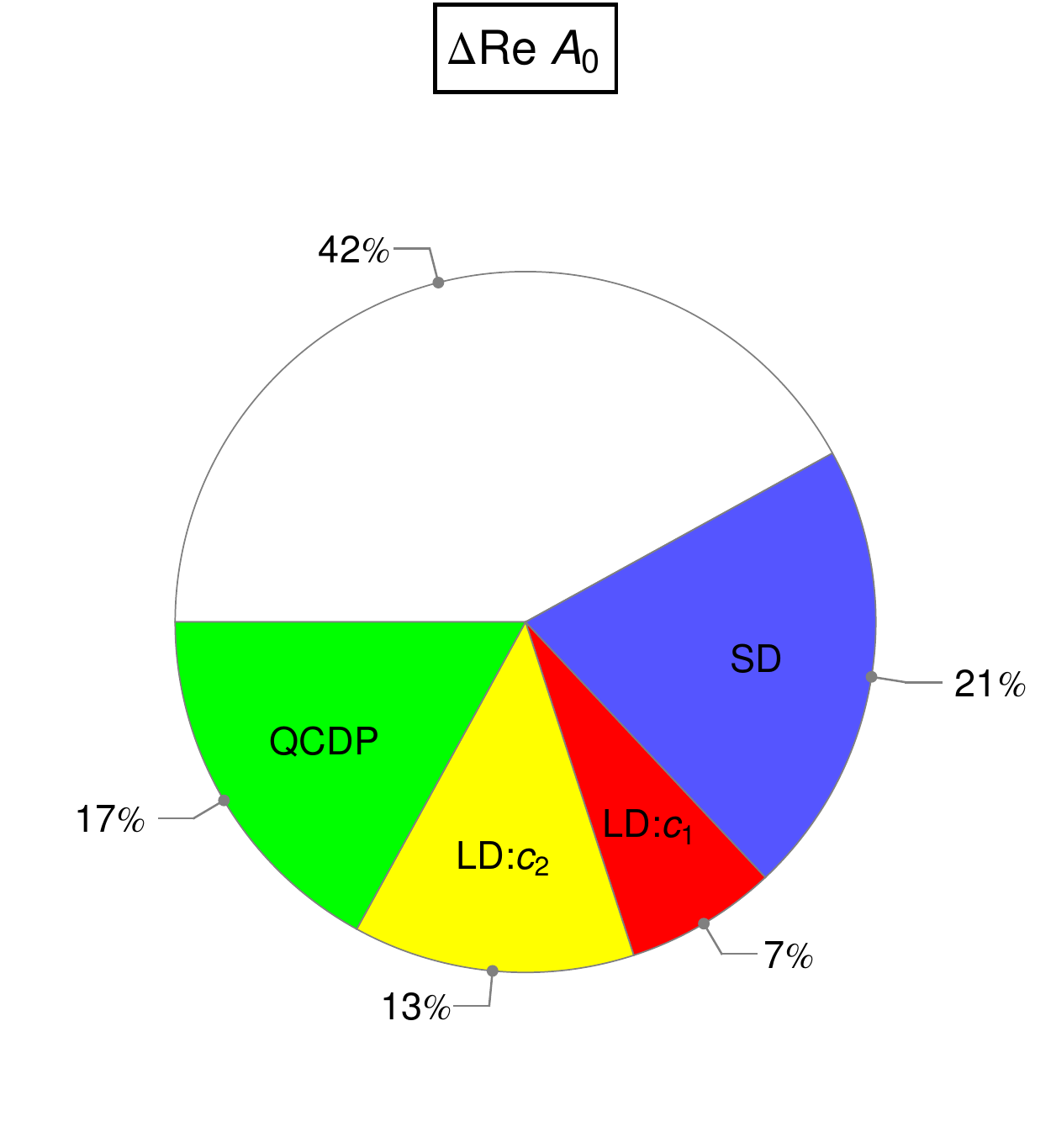}
\end{center}
\vskip-0.8cm
\caption{Budgets for ${\rm Re}A_2$ (left) and   $\Delta{\rm Re}A_0$ (right) 
summarizing the size of different suppression mechanisms of 
${\rm Re}A_2$ and enhancement mechanisms of ${\rm Re}A_0$, denoted here 
by $\Delta{\rm Re}A_0$, for the matching scale $\mu=M=0.8\gev$. SD stands for 
quark-gluon evolution and LD for meson evolution. In the case of 
 $\Delta{\rm Re}A_0$ we decompose LD into contributions coming from $c_1$ and $c_2$. 
QCDP stands for $Q_6$ contribution. See text for detail explanations.}
\label{fig:pies}
\end{figure}

Finally, in Fig.~\ref{fig:pies} we show budgets for ${\rm Re}A_2$ (left) and ${\rm Re}A_0$ (right) that summarize
the size of different suppression mechanisms of 
${\rm Re}A_2$ and enhancement mechanisms of ${\rm Re}A_0$\footnote{ We thank Jennifer Girrbach for providing these plots.}. SD stands for 
quark-gluon evolution and LD for meson evolution. In the case of 
 ${\rm Re}A_0$ we decompose LD into contributions coming from $c_1$ and $c_2$. 
QCDP stands for $Q_6$ contribution. We set the matching scale at $\mu=0.8\gev$.

In the case of ${\rm Re}A_2$ the division into SD and LD contributions is 
straightforward as seen in (\ref{BBG2}). We find then that $52\%$ of suppression  of  ${\rm Re}A_2$ comes from SD ({\it violet}) and $48\%$ from LD ({\it red}).
The colour coding expresses the ultraviolet and infrared 
character of the two contributions, respectively.

The case  of  ${\rm Re}A_0$ is more complicated in view of 
the fact that penguin contributions are present, the LD contributions involve 
the coefficients $c_1$ and $c_2$ and finally, as we have seen, we are not able 
to explain fully the missing   $\Delta{\rm Re}A_0=23.4\times 10^{-8}\gev$ 
relative to the large $N$ limit. We will normalize different contributions 
in the budget to this additive contribution required by the data. Following 
\cite{Bardeen:1986uz} we will assume that due to incomplete GIM mechanism 
above $m_c$ QCD penguin contributions are enhanced at $\mu=0.8\gev$ by a factor 
of two so that $|z_6|\bsi=0.08$. Using the results in Table~\ref{tab:final} 
we find then that $42\%$ of the missing shift in  ${\rm Re}A_0$ remains 
presently unexplained and we present it in {\it white}. In this normalization 
QCD penguin contribution amounts to $17\%$ and corresponds to the {\it green} 
area with the colour chosen to express the character of this particular 
contribution.

The division between SD and LD current-current contributions to 
 $\Delta{\rm Re}A_0$  is complicated 
by the fact, as seen in (\ref{BBG1}), that in contrast to ${\rm Re}A_2$ 
the coefficients of $z_1$ and $z_2$ involve different LD factors. Therefore 
just swiching off LD part or SD part can only teach us about relative 
importance of these two contributions but their sum will miss by a factor of 
$1.4$ the total contribution from octet enhancement that one obtains when these 
two contributions are simultaneously at work. Correcting for this 
factor we finally find that the $41\%$ contribution from octet enhancement of
 $\Delta{\rm Re}A_0$
is, like for ${\rm Re}A_2$, almost equally distributed between 
these two contributions: SD ($21\%$)  and LD ($20\%$). 
 In order to stress the importance of 
the mixing of $Q_2$ and $Q_6$ operators we divide the LD contribution into 
two parts so that the effect of $c_2$ in enhancing ${\rm Re}A_0$ is roughly twice as large as the one  of $c_1$.
The part of LD related to $c_1$ is again in {\it red} but $c_2$ area representing the 
mixing of $Q_2$ and $Q_6$ or equivalently mixing of {\it red} and {\it green} is
consequently {\it yellow}.

{Finally, we would like to refer to the analysis in \cite{Bijnens:1998ee} 
    which was done in the spirit of our approach except that for the 
  low energy meson contributions an Extended Nambu-Jona-Lasinio model has 
  been used resulting in  differences in the matching 
  between long distance and short distance contributions. Also these authors 
 find sizable enhancement of ${\rm Re}A_0$  and suppression 
of ${\rm Re}A_2$ but various uncertainties in their model 
allow them only to quote the range
$15\le R \le 40$. The large value of $R$ originates in a small value of 
 ${\rm Re}A_2$ which is stronger suppressed than required and is typically 
by $30\%$ below its experimental value.}

\section{Comparison with Lattice Results}\label{Lattice}
We will now compare our results with 
the results on
${\rm Re}A_0$  and ${\rm Re}A_2$ 
 from the RBC-UKQCD collaboration \cite{Boyle:2012ys,Blum:2011pu,Blum:2011ng,Blum:2012uk}. As the normalization 
of ${\rm Re}A_0$ and ${\rm Re}A_2$  in the latter papers differs from ours, 
we first have to define
\be\label{NORM}
{\rm Re}A_0=\sqrt{\frac{2}{3}}({\rm Re}A_0)_L, \qquad 
{\rm Re}A_2=\sqrt{\frac{2}{3}}({\rm Re}A_2)_L,
\ee
where subscript $L$ refers to the amplitudes in \cite{Boyle:2012ys}.
The latter are given in terms of contractions $\circled{1}$ and $\circled{2}$ 
in Fig.~1 of that paper that correspond to the diagrams (a) and (b) 
in our Fig.~\ref{Fig:1}, respectively. One has to be careful in this comparison as in 
 \cite{Boyle:2012ys} the Fierz transformed form of $Q_1$ relative to the one 
in (\ref{current}) is used. Basically, $Q_2$ contributes to $K^0\to\pi^+\pi^-$
and $K^0\to\pi^0\pi^0$ through contractions $\circled{1}$ and $\circled{2}$, 
respectively, while in the case of $Q_1$ the role of contractions is 
interchanged. With this information, the diagrams  (c) and (d) in Fig.~\ref{Fig:1}  are 
automatically included in the results for the amplitudes which read 
 \cite{Boyle:2012ys}:
\be\label{L1}
({\rm Re}A_0)_L=\frac{G_F}{\sqrt{2}}V_{ud}V_{us}^*\left(\frac{1}{\sqrt{3}}\right)
\left[z_1\left(2~\circled{2}-\circled{1}\right)+z_2\left(2~\circled{1}-\circled{2}\right)\right],
\ee

\be\label{L2}
({\rm Re}A_2)_L=\frac{G_F}{\sqrt{2}}V_{ud}V_{us}^*\sqrt{\frac{2}{3}}
(z_1+z_2)\left(\circled{1}+\circled{2}\right)~.
\ee

Before comparing with our results let us find out what values of $\circled{1}$ and 
$\circled{2}$ at $\mu=2.15\gev$, used in \cite{Boyle:2012ys}, would simultaneously reproduce the data for both amplitudes. With the  NDR-${\rm \overline{MS}}$ 
values $z_1=-0.287$ and $z_2=1.133$ at  $\mu=2.15\gev$, we find
\be\label{CON1}
\circled{1}=0.0791\gev^3,\quad        \circled{2}= -0.0673 \gev^3,\quad 
\circled{2}=-0.85~ \circled{1}.
\ee

It should be emphasized that these results apply to the  NDR-${\rm \overline{MS}}$ scheme and, as the contractions represent the matrix elements, they must 
be both scheme and scale dependent.

Now in  \cite{Boyle:2012ys} $\circled{2}\approx -0.7~ \circled{1}$ is found. 
However it should be stressed that this numerical result is not in 
the  NDR-${\rm \overline{MS}}$ scheme but in the lattice scheme used 
there \footnote{Chris Sachrajda, private communication.}.  The relative sign of 
these two contractions found in \cite{Boyle:2012ys} is an important result 
and agrees with the sign  we would obtain using the same language, as 
discussed in more detail below. 

We note also that in our normalization the lattice result for  ${\rm Re}A_2$ 
in  \cite{Blum:2012uk} reads:
\be\label{A2Lattice}
{\rm Re}A_2= (1.13\pm0.21)\times 10^{-8}~\gev.
\ee
The error is dominated by systematics. This result is 
in agreement with the data and, within uncertainties, with our results for ${\rm Re}A_2$ in Tables~\ref{tab:DIRule} and \ref{tab:final}. In fact, though obtained using a different approach, we find it remarkable 
that the central value in (\ref{A2Lattice}) differs from our central 
value in (\ref{Step2}) by only $6\%$. This is still another support 
for the dual picture of QCD.

Not having the Wilson coefficients $z_1$ and $z_2$ in the lattice scheme, but 
expecting that in the future all lattice results will be quoted in the 
NDR-${\rm \overline{MS}}$ scheme used by phenomenologists,
we may nevertheless investigate how the result for ${\rm Re}A_0$ depends 
on the ratio of  these two contractions in the latter scheme assuming the data  for  ${\rm Re}A_2$. 
Defining then the $K$ factor by 
\be\label{CON2}
\circled{2}=-K~\circled{1} 
\ee
we show in Table~\ref{tab:K} the results for the two contractions, ${\rm Re}A_0$ and $R$ for different values of $K$. As we use $z_{1,2}$ in the NDR-${\rm \overline{MS}}$ scheme for $\mu=2.15\gev$, these results apply only to this scheme and this scale.

We observe that the final results for the quantities in Table~\ref{tab:K} strongly 
depend on the value of $K$ and for $K\approx 0.7$, the ratio $R$ is in the 
ballpark of the ratio found in  \cite{Boyle:2012ys}, even if a different 
scheme is used there. Yet, in view of comments 
made above and the fact  that the lattice result for ${\rm Re}A_0$ 
corresponds to non-physical kinematics this comparison is only on a qualitative level. Still the message 
is clear. If the ratio $K$ in the NDR-${\rm \overline{MS}}$ scheme will be 
found significantly smaller than $K=0.85$ and agreement with 
the data on ${\rm Re}A_2$ will be imposed,  a satisfactory description of the 
data on  ${\rm Re}A_0$,  even at scales $\mu=2-3\gev$, will not be possible with $\circled{1}$ and $\circled{2}$ only. The rescue could come then from 
other contractions that involve QCD penguin contributions. These contributions 
are presently estimated in \cite{Boyle:2012ys} to be very small. But the situation may change when the calculations are performed at physical kinematics.

\begin{table}[!tb]
\centering
\begin{tabular}{|c|c|c|c|c|c|c|}
\hline
 $K$  &  $0.50$ & $0.60$ & $ 0.70$&  $0.80$ & $0.85$ & $0.90$\\
\hline
\hline
\parbox[0pt][1.6em][c]{0cm}{}$\circled{1}[\gev^3]$  &  $0.0237$  &$0.0296$& $0.0395$ &$0.0593$ & $0.0791$ & $0.119$ \\
\parbox[0pt][1.6em][c]{0cm}{}$\circled{2}[\gev^3]$  &  $-0.0119$  &$-0.0178$& $-0.0277$ &$-0.0474$ & $-0.0673 $ & $-0.107$ \\
\parbox[0pt][1.6em][c]{0cm}{}$10^{8}{\rm Re} A_0[\gev]$  &  $6.9$  &$9.0$& $12.6$ &$19.8$ & $27.0$  & $41.3$\\
\parbox[0pt][1.6em][c]{0cm}{}$R$&  $5.7$& $7.5$ &$10.4$&$16.4$& $22.3$ & $34.2$   \\
\hline
\end{tabular}
\caption{The two contractions  in 
the NDR-${\rm \overline{MS}}$ scheme for $\mu=2.15\gev$
and resulting 
${\rm Re} A_0$   and $R$  for different values of $K$ defined in (\ref{CON2}) 
assuming ${\rm Re}A_2$ to agree with data.}
\label{tab:K}~\\[-2mm]\hrule
\end{table}

Comparing the expressions (\ref{L1}) and (\ref{L2}) with our results in (\ref{BBG1}) and (\ref{BBG2}) and taking into account different normalization we 
can express the contractions $\circled{1}$ and $\circled{2}$ in terms of $X_F$ and $c_1$. To this end we have to set $c_2=0$ and drop penguin contributions. 
We find then 
\be\label{dic}
\circled{1}= \frac{X_F}{\sqrt{2}}, \qquad \circled{2}= - c_1 \frac{X_F}{\sqrt{2}}, \qquad K=c_1.
\ee
It should be remembered that contractions and also $K=c_1$ are scheme  and scale dependent and the ones 
given here are in the ${\rm \overline{MOM}}$ scheme. However,
already this  result  offers 
the explanation of the {\it positive} sign of $\circled{1}$ and of  {\it negative} sign of $\circled{2}$ found in  \cite{Boyle:2012ys}. 
 In particular, the latter sign follows in our approach from 
the proper matching of the anomalous dimension matrices in the meson and 
quark-gluon pictures of QCD \footnote{As a side remark let us note that within 
VIA $K=c_1=-1/3$ which is at variance not only with our results but also with 
the findings in \cite{Boyle:2012ys}.}. Therefore the  result obtained
 in  \cite{Boyle:2012ys} is an important support for our dual QCD 
approach to weak decays, in particular as the lattice calculations will 
eventually provide much more precise results than can be obtained in our 
analytic approach. Even if with $X_F=0.0298~\gev^3$ the values of the contractions in (\ref{dic}) appear 
at first sight to be much smaller than the ones collected in Table~\ref{tab:K},  it can be demonstrated  that they are fully compatible with the dynamics at scales $\ord(2\gev)$.

Indeed,  the authors of  \cite{Boyle:2012ys} work at  $\mu=2.15\gev$ and 
we at $M\approx 0.8\gev$. Therefore  our $K$ factor must be different than the one in 
lattice calculations. It must be smaller and, as seen in Table~\ref{tab:Results}, this is indeed the case. Therefore the numerical comparison of the results of \cite{Boyle:2012ys}
with ours must also involve the Wilson coefficients $z_i$. 
The fact that our approach and lattice approach predict similar values for 
 ${\rm Re}A_2$ implies the compatibility 
of both approaches as far as $\Delta I=3/2$ transitions are concerned.

The comparison of both approaches in the case of  ${\rm Re}A_0$ is more 
difficult because in our approach the QCD penguin contributions cannot 
be neglected. Moreover, in our approach the mixing of $Q_2$ operator with $Q_6$ operator represented by $c_2$ constitutes a significant part of the 
enhancement of  ${\rm Re}A_0$  in the current-current sector. We have emphasized it in previous sections and in Fig.~\ref{fig:pies}. On the basis 
of the formulae (\ref{X1}) and (\ref{X3}) we expect that the latter effects 
are present in the hadronic matrix elements of the operator $Q_2$ evaluated 
at the lattice scales.

In this context it is interesting to note that in the strict large $N$ limit 
\be
\circled{1}=0.0210~\gev^3, \qquad \circled{2}=0, \qquad (\mu\approx 0)
\ee
which drastically differs from the values of contractions in Table~\ref{tab:K} 
for $K\ge 0.6$ that correspond to $\mu=2.15\gev$. Yet the fast meson evolution and the presence of significant 
QCD penguin contributions, both through their diagonal evolution and 
mixing with $Q_2$ operator, allows us, as seen in Tables~\ref{tab:final} and 
\ref{tab:K}, to obtain values of ${\rm Re}A_0$ that with the contractions considered in \cite{Boyle:2012ys} can only be obtained for $K$ as large as 
$K\approx 0.75$ within the NDR-${\rm \overline{MS}}$ scheme.

This discussion shows, that at least at a semi-quantitative level, the 
recent lattice results can be interpreted within 
 the dual representation of QCD as a theory of weakly interacting mesons 
for large $N$. A more detailed comparison will only possible when lattice 
results for  ${\rm Re}A_0$ with physical kinematics will be available.

\section{$K_L-K_S$ Mass Difference}\label{sec:9}
\subsection{Preliminaries}
We begin our discussion by summarizing the status of short distance 
contributions to $\Delta M_K$ within the SM. 
For that purpose we decompose it as follows:
\be\label{LD1}
\Delta M_K = (\Delta M_K)_{cc}+(\Delta M_K)_{ct}+(\Delta M_K)_{tt}+(\Delta M_K)_{\rm LD},
\ee 
with the first three short distance contributions obtained from usual 
box diagrams and the last term standing for long distance contributions.
The second and third term contributing at most $1\%$ to $\Delta M_K$ 
\cite{Brod:2011ty,Buras:2013raa} will be neglected in what follows.
 For the dominant 
contribution we have 
\be\label{LD2}
(\Delta M_K)_{cc}=
\frac{G_F^2}{3\pi^2} (V_{ud}V^\ast_{us})^2 F_K^2 \hat B_K m_K \eta_{cc} m_c^2(m_c)~.
\ee
The QCD factor $\eta_{cc}$ including NLO \cite{Herrlich:1993yv}  and NNLO \cite{Brod:2011ty}  QCD corrections is unfortunately subject to very large uncertainties 
\be\label{Brod}
\eta_{cc}=1.87(76)
\ee
so that  \cite{Brod:2011ty}
\be
(\Delta M_K)_{cc}= (3.1\pm 1.2) 10^{-15}\gev=(0.89\pm 0.34)(\Delta M_K)_{\rm exp}
\ee
with the experimental value given in (\ref{DMK}). We conclude therefore that 
extracting $(\Delta M_K)_{\rm LD}$ from the data on the basis of this calculation is impossible as this would imply the range of values between $45\%$ to $-21\%$ of the measured value. As we will demonstrate below, from our approach 
 $(\Delta M_K)_{\rm LD}$ is known much better and this invited the authors 
of \cite{Buras:2013raa} to use this result for the extraction of $\eta_{cc}$ 
from $(\Delta M_K)_{\rm exp}$. In this manner the uncertainty in 
the evaluation of $\varepsilon_K$ could be reduced.
\boldmath
\subsection{$(\Delta M_K)_{\rm LD}$ in the the strict Large $N$ Limit}
\unboldmath
We have seen that   the large $N$ value for $\hat B_K$ is supported by the latest lattice results. So, we feel rather confident about calculating the $K_L-K_S$ mass difference within the same approximation \cite{Gerard:1990dx,Bijnens:1990mz}.   

In order to get some feeling for the size of effects, we calculate first 
$(\Delta M_K)_{cc}$ in the strict large $N$ limit. In this case $\hat B_K=3/4$ 
but in addition $\eta_{cc}=1$ in (\ref{LD2}).  Yet for the very low values of scales used 
for the evaluation on $\hat B_K$ we cannot use $m_c(m_c)$ but rather its 
constituent mass $m_c=1.5\pm 0.1\gev$. This rough estimate results in $(66\pm 9)\%$ 
of the measured value attributed to short distance part and $+(34\pm 9)\%$ to the 
LD contribution. The important message from this simple exercise is the 
positivity of $(\Delta M_K)_{\rm LD}$. Yet, we would like to provide a better 
estimate.

Applying this strategy but not using the constituent charm quark mass,
it is quite convenient to parametrize the full $\Delta M_K$ as follows  ($B_K = 3/4$):
\be
\Delta M_{K} = \frac{G_F^2}{4\pi^2} (V_{ud}V^\ast_{us})^2 F^2_K m_K M^2_\Delta = (10^{-15} \ \textrm{GeV}^{-1}) M^2_\Delta.
\ee

From the experimental value (\ref{DMK}) of this mass splitting,
we easily  extract
\be
M^{\exp}_\Delta = 1.87 \ \textrm{GeV}.
\ee
In the effective Fermi theory, such a scale has been associated with the mass of some new degree of freedom to appear in the UV completion. First misidentified as the mass of a hypothetical $W$ weak boson, this $\Delta$S = 2 scale has then been eventually linked (with the help of the GIM mechanism) to the mass of a yet-to-be-discovered charm quark \cite{Gaillard:1974hs}. Working again in the $m^2_\pi = 0$ limit, let us estimate $M_\Delta$ in the large $N$ limit.

A straightforward calculation of the standard box-diagram involving only virtual charm or (and) up quarks gives then
\be\label{GIM}
M^2_\Delta (SD) = m^2_c - M^2 + m^2_K \ln (m^2_c/M^2) - (5/6) m^2_K +  \mathcal{O} (m^4_K/M^2),
\ee
if $M$ is the IR cut-off for the high $W$-momenta:
\be
M^2 < q^2_W < m^2_c.
\ee
In (\ref{GIM}), the relative sign between the first two quadratic terms results from the GIM mechanism at work $(m_u = 0)$ while the third logarithmic one arises when keeping the external momentum for the strange quarks $(m_d = 0)$.

At long-distance, the $K$ and $\pi$ one-loops generated by $Q_2\otimes Q_2$ give
\be\label{Gnew}
M^2_\Delta (LD) = (7/4) M^2 - (3/4) m^2_K \ln (M^2/m^2_K) + (11/24) m^2_K +  \mathcal{O} (m^4_K/M^2),
\ee
if $M$ is the UV cut-off for the low $W$-momenta:
\be
0 < q^2_W < M^2.
\ee
With this unambiguous identification of the momentum across the SD-LD frontier, we can consistently impose the $M$-independent condition
\be
\partial/\partial M^2 [M^2_\Delta (SD)+M^2_\Delta(LD)]=0
\ee
to get an optimal matching scale remarkably close to the light vector meson mass, namely
\be
M = \sqrt{\frac{7}{3}} m_K \approx m_V\approx 0.8\gev.
\ee
If we vary the cut-off around this natural matching scale (say, 0.5 GeV $<$ $M$ $<$ 1.0~GeV), the LD contribution relative to the measured $\Delta M_K$ mass difference turns out to be (30 $\pm$ 15)\% in a remarkable agreement with our 
previous estimate.    But within  our dual picture of QCD we always have to combine the LD contribution with its complementary, namely the SD one, to get any observable. Doing so with (\ref{GIM}) and (\ref{Gnew}),
we now observe a remarkable stability with respect to variations of $M$ in the same energy range: 
\be
\Delta M_{K} (SD+LD)=(0.80 \pm 0.10)( \Delta M_{K})_{\rm exp}.
\ee
In fact, the main uncertainty in this large $N$ estimate of the $K_L-K_S$ mass difference arises from the charm quark (constituent) mass taken here to be $m_c = (1.5 \pm 0.1)$~GeV.

\subsection{Non-Leading Corrections}
In the $1/N$ expansion, leading and sub-leading contributions to $\Delta M_{K}$ correspond to the same topologies as for the $B_K$, once the fictitious color singlet boson is replaced by two physical $W'$s. Consequently, one might expect the $1/N$ corrections to the $K_L-K_S$ mass difference to be negative and thereby  modifying our previous estimate. As we will show now this is indeed the case for the LD ($\pi$, $\eta$ and $\eta'$) pole contributions generated this time by 
 $Q_1\otimes Q_1$. However, we already know from $B_K$ how a partial 1/N estimate can misrepresent the physical world.

Our simple analytical approach can be extended to the full nonet of pseudo-scalars ($\eta_0$ included) to disentangle the QCD penguin operator $Q_6$ from $Q_2- Q_1$:
\be
Q_6(0) = - (r^2/\Lambda^2_\chi) (Q_2-Q_1+Q_3)(0)
\ee
with the new current-current operator
\be
Q_3 = 4 (\overline{s_L} \gamma^\mu  {d_L})(\overline{q_L} \gamma_\mu  {q_L})
\ee
proportional to $\partial_\mu\eta_0$. 
As a result, it can easily be applied to other observables somehow related to the  empirical $\Delta I = 1/2$ rule, such as  radiative $K$-decay rates \cite{Gerard:2005yk} or
the  $\ord(G^2_F \varepsilon')$ weakly-induced strong $\theta$ parameter \cite{Gerard:2012ud}.
 In the same manner the $1/N$-suppressed $(\pi,\eta,\eta')$ pole contribution to the 
$\Delta M_K$ are found to be \cite{Gerard:2005yk}
\be\label{pole}
\Delta M_{K}({\rm pole})\approx -0.3 (\Delta M_K)_{\rm exp},
\ee
canceling significantly the leading order estimate. Our final estimate of LD 
contributions to $\Delta M_K$ within our approach including estimates of 
leading and next to leading corrections gives then
\be\label{FinalLD}
(\Delta M_{K})_{\rm LD}\approx (0.2\pm 0.1) (\Delta M_K)_{\rm exp}.
\ee
This result is consistent with the analysis of $(\Delta M_K)_{\rm LD}$   in the context of the calculation of long distance
effects in $\varepsilon_K$ \cite{Buras:2010pza}.
Using it in (\ref{LD1}) and (\ref{LD2}) and assuming no new physics 
contributions to $\Delta M_{K}$, one extracts $\eta_{cc}$ from the data to 
be \cite{Buras:2013raa}
\be
\eta_{cc}=1.7\pm 0.2
\ee
with an error almost four times smaller than the error in the direct 
calculation in (\ref{Brod}). It should be emphasized that this value 
should not be confused with $\eta_{cc}=1$ used in our exercise before as 
in this extraction $m_c(m_c)=1.28\gev$ has been used in order to compare 
with the result in (\ref{Brod}). If $m_c=1.5\gev$ was used instead, we 
would find $\eta_{cc}=1.23\pm 0.15$, fully compatible with $\eta_{cc}=1$. 
We note that for the computation of charm contribution to $\varepsilon_K$ 
only the product $\eta_{cc}m_c^2$ enters and if $\eta_{cc}$ is extracted 
from experimental value of $\Delta M_{K}$ it is immaterial which of these 
two values of $m_c$ are used.

Needless to say, we are aware of the fact that our estimate in (\ref{FinalLD})
requires more detailed investigations and in 
particular future confirmation from lattice 
simulations. Presently, no reliable result on $(\Delta M_K)_{\rm LD}$ from lattice is 
available but an important progress towards its evaluation has been 
made in \cite{Christ:2012se}. This first result seems to indicate that 
$(\Delta M_K)_{\rm LD}$ could be larger than expected by us. We are therefore looking forward to more precise evaluation of this important quantity from the lattice in order 
to see whether also in this case large $N$ approach passed another test or not.

\section{Conclusions}\label{sec:8}
Motivated by the recent advances in the computation of non-perturbative 
parameters in the Kaon system by several lattice collaborations \cite{Tarantino:2012mq,Sachrajda:2013fxa,Christ:2013lxa}, in particular the RBC-UKQCD  collaboration, 
we have reviewed our results obtained in the 1980s within the dual representation of QCD as a theory of weakly interacting mesons for large $N$. This includes 
in particular:
\begin{itemize}
\item
The parameter $\hat B_K$,
\item
The isospin amplitudes ${\rm Re} A_0$ and ${\rm Re} A_2$,
\item
$K_L-K_S$ mass difference.
\end{itemize}

It is remarkable that the recent lattice QCD results using dynamical fermions 
confirm our finding of 1980s that $\hat B_K$ is very close to its large $N$ 
value. Relative to our first paper on $\hat B_K$ \cite{Bardeen:1987vg}, 
where only pseudoscalar meson contributions have been taken into account, 
the inclusion of vector meson contributions, already advocated by one of us in 
 \cite{Gerard:1988it,Gerard:1990dx}, decreased significantly the left-over 
scale dependence of $\hat B_K$ bringing it very close the its large $N$ 
value of $3/4$. The numerical confirmation of this result by a number of 
lattice groups gives support for our work of 1980s.
The smallness of $1/N$ corrections to the large $N$ value $\hat B_K=3/4$ results, within our approach, from an approximate cancellation between the  pseudoscalar and vector meson one-loop contributions. This is clearly 
demonstrated in Table~\ref{tab:BKResults}.

Concerning $\Delta I=1/2$ rule our physical explanation, stated already 
in the abstract and discussed in detail in Section~\ref{sec:5a}, is based 
on the evolution from high energy scales down to very low energy scales 
at which factorization of the hadronic matrix elements into products 
of current matrix elements in the case of current-current operators and quark densities in the case of QCD penguin operators is recovered. As the long but 
slow
quark-gluon evolution and short but fast meson evolution involve different degrees of freedom the matching around $\ord(1\gev)$ scale is more challenging than 
in lattice QCD which works with quarks and gluons only. Yet, as we have shown,  
when vector meson contributions are included and the Wilson coefficients are calculated in the ${\rm \overline{MOM}}$ scheme the matching is very good  in the 
case of ${\rm Re} A_0$ but also satisfactory for  ${\rm Re} A_2$ which is found 
 close to its experimental value and also to its lattice value. We expect that 
the inclusion of heavier resonances and going beyond the chiral limit estimate 
of vector meson contributions will further bring the theory
closer to the data.

As seen in Table~\ref{tab:DIRule}, the current-current operators alone can 
at scales considered by us explain roughly $60\%$ of the $\Delta I=1/2$ rule. 
As pointed out by us in  \cite{Bardeen:1986vz}, this should be considered as the dominant mechanism of the $\Delta I=1/2$ rule as it suppresses  ${\rm Re} A_2$
amplitude and enhances  ${\rm Re} A_0$. It should be emphasized that the 
quark-gluon evolution with the present value of $\alpha_s$ is insufficient 
to suppress  ${\rm Re} A_2$ in order to reproduce the data. Additional 
suppression is necessary from hadronic matrix elements. In our approach this 
is achieved through fast meson evolution, which while suppressing ${\rm Re} A_2$
enhances further ${\rm Re} A_0$. The recent findings by the  RBC-UKQCD lattice collaboration confirm this picture in a spectacular manner in the case of 
${\rm Re} A_2$, but also the enhancement of ${\rm Re} A_0$ in 
lattice simulations is very interesting. While the latter approach obtains 
presently $R\approx 11$, the results for  ${\rm Re} A_0$ are still obtained 
using non-physical kinematics and improving on this in the future 
should enhance $R$ towards its experimental value.

Yet, at the scales we are working, QCD penguins provide a significant 
contribution to  ${\rm Re} A_0$, in particular as the value of the strange quark mass 
decreased  relative to our analysis in 1986.  We find then $R\approx 16.0\pm 1.5$ that depends on the size of incomplete GIM mechanism that deserves further 
study in the ${\rm \overline{MOM}}$ scheme together with $1/N$ corrections to the hadronic 
matrix elements of $Q_6$. These effects and inclusion of  higher mass
resonances could provide the explanation of the missing $30\%$ in ${\rm Re} A_0$.  The present budgets of different mechanisms suppressing ${\rm Re} A_2$ and 
enhancing  ${\rm Re} A_0$ in our approach are summarized in Fig.~\ref{fig:pies}.

In the case of lattice calculations normalized around  $2~\gev$, explicit QCD penguin contributions to  ${\rm Re} A_0$  are much smaller as the GIM suppression is still rather effective at these scales.
The significant contribution of QCD penguins should then 
be found in the enhanced matrix elements of current-current operators, in 
particular $Q_2$ operator. In our approach this corresponds to the 
increased value of the coefficient $c_2$, which as seen in Table~\ref{tab:Results}, increases with increased value of $M$. This increase, as seen in 
(\ref{BBG1}), enhances  ${\rm Re} A_0$  in addition to the enhancement through 
$c_1$.

From the point of view of our approach the  RBC-UKQCD lattice collaboration clearly identified the effects in both amplitudes coming from the enhanced 
value of $c_1$. The next step would be to separate the enhancement of 
${\rm Re} A_0$ through $c_1$ from the one through $c_2$. This would be signalled  by an enhanced matrix elements of the $Q_2$ operator in $K^0\to\pi^+\pi^-$ 
and $K^0\to\pi^0\pi^0$ decays. It should also be  investigated whether 
the role of QCD penguin operator $Q_6$ at these higher scales is indeed as small as presently implied by lattice results. It would also be interesting 
to perform lattice calculations of hadronic matrix elements at several values 
of $\mu$ including those considered in our paper in order to verify 
meson evolution of hadronic matrix elements more precisely than can be done in our approach.

While our analytic approach allowed us to identify the  dynamics behind the 
observed $\Delta I=1/2$ rule, the precision calculations of  ${\rm Re} A_0$ and 
 ${\rm Re} A_2$ can only be obtained from lattice QCD although it will 
take some time before uncertainties in these amplitudes will be reduced 
down to $10\%$ level. Whether the lattice approach will be able, on its own, 
to provide the physical explanation of the dynamics behind the $\Delta I=1/2$ 
rule remains to be seen. It would also be important to make further efforts in the context of realistic AdS/QCD descriptions of the 1/N expansion as this should allow more a precise interpolation of the meson amplitudes to scales explored by the lattice community.  In this manner, the comparison of the 1/N expansion with the unquenched lattice results could be made more explicit.  It would also allow a closer look at the upper bound on $\hat B_K$ at these higher energy scales.

In summary, it is quite encouraging that our simple analytic framework improved 
by the inclusion of vector mesons and proper matching to short distance 
Wilson coefficients yields consistent results in good agreement with the 
data. Simultaneously, it provides a simple picture of the dynamics behind 
the $\Delta I=1/2$ rule which as the basis has the main property of QCD: 
asymptotic freedom and the related evolutions of weak matrix elements 
which at long distance scales can be performed in the dual representation 
of QCD as a theory of weakly interacting mesons for large $N$.

\section*{ Acknowledgements}
AJB would like to thank Gino Isidori and Heiri Leutwyler for very encouraging comments on the first version of the paper and Jure Drobnak and Robert Ziegler for checking numerically the values of the Wilson coefficients $z_i$.
We would like to thank Jennifer Girrbach, Chris Sachrajda and Amarjit Soni for discussions and Nuria Carrasco, Luca Silvestrini  and Vittorio Lubicz for E-mail exchanges.
This research was financially supported by the ERC Advanced Grant project ``FLAVOUR'' (267104) and the Belgian IAP Program BELSPO P7/37. 
It was also  partially supported by the DFG cluster
of excellence ``Origin and Structure of the Universe''.
Fermilab is operated by Fermi Research Alliance, LLC under Contract No. DE-AC02- 07CH11359 with the United States Department of Energy.

\bibliographystyle{JHEP}
\bibliography{NREF.bib}
\end{document}